\documentclass[11pt,a4paper]{article}

\usepackage[utf8]{inputenc}
\usepackage{amsmath, amssymb, amsthm}
\usepackage{graphicx}
\usepackage{subcaption}
\usepackage{geometry}
\usepackage{setspace}
\usepackage{booktabs}
\usepackage{xcolor}
\usepackage{multirow}
\usepackage{natbib}
\usepackage{tabularx}
\usepackage{enumitem}
\usepackage{arydshln}

\RequirePackage[colorlinks=true,allcolors=teal]{hyperref}

\newcommand{\highlightitem}[1]{%
    \par\noindent
    \hangindent=1.6em
    \hangafter=1
    \makebox[1.6em][l]{$\bullet$}#1\par\vspace{0.6em}
}

\title{Power Play: How Structural Shocks Drive \\ European Electricity Markets}

\author{
Francesco Ravazzolo\textsuperscript{1,2},
Luca Rossini\textsuperscript{3,4},
and Andrea Viselli\textsuperscript{2}
\thanks{
The authors gratefully thank the participants at the IAEE Annual conference 2026 in Lisbon, SiDE Summer School in Bertinoro, 2nd Conference on Applied Macroeconomics in Rome on October 8–9, 2026, at the seminars at University of Milan, and Technische Universitat Dortmund, for their valuable comments and discussions. Francesco Ravazzolo acknowledges financial support from the Centre for Advanced Study (CAS) at the Norwegian Academy of Science and Letters. Andrea Viselli acknowledges financial support from the SURE-AI Centre (Grant No. 357482), funded by the Research Council of Norway.
Correspondence to: Andrea Viselli, BI Norwegian Business School, Nydalsveien 37, 0484 Oslo, Norway.  
E-mail: \url{andrea.viselli@bi.no}.
} \\[1em]
{\small\textsuperscript{1}Free University of Bozen, Italy} \\
{\small\textsuperscript{2}BI Norwegian Business School, Norway} \\
{\small\textsuperscript{3}University of Milan, Italy} \\
{\small\textsuperscript{4}Fondazione Eni Enrico Mattei, Italy}
}

\date{\today}

\begin{document}
\maketitle

\begin{abstract}
We identify the structural drivers of day-ahead electricity prices in Germany and Italy during the European energy crisis and its aftermath. We estimate a daily, recursively identified SVAR from January 2021 to {September} 2025, combining electricity prices, load and renewable-generation forecasts, fossil-fuel prices, and EU ETS carbon prices. The identification is motivated by the timing of day-ahead electricity auctions and by the merit-order mechanism. Renewable-supply shocks lower electricity prices on impact, especially wind shocks in Germany, but their effects are relatively short-lived. Fossil-fuel shocks, and gas-price shocks in particular, become more important at longer horizons and explain a large share of the 2022 electricity-price surge. The cross-country comparison reveals marked heterogeneity: Germany displays a stronger renewable-supply channel, whereas Italy remains more exposed to gas-price and electricity-market-specific shocks. The results indicate that reducing fossil-fuel exposure and increasing system flexibility are central to limiting electricity-price volatility in renewable-intensive power systems.


\end{abstract}

\newpage
\vspace{1cm}
\begingroup
\renewcommand{\abstractname}{Highlights}
\begin{abstract}
\highlightitem{Separate crisis and post-crisis decompositions show that the end of the energy crisis did not imply a uniform decoupling from gas: Italy remains strongly gas-exposed, while Germany shifts more toward wind and electricity-market-specific shocks.}

\highlightitem{Renewable-supply shocks lower electricity prices on impact, but their effects are less persistent than fossil-fuel shocks; this suggests that batteries, storage, and system flexibility are needed to make renewable price benefits more durable.}

\highlightitem{Gas shocks explain a large share of the 2022 electricity-price surge and dominate long-horizon dynamics, supporting policies aimed at reducing fossil-fuel exposure rather than redesigning short-run marginal pricing per se.}
\end{abstract}
\endgroup

\newpage
\doublespacing

\section{Introduction}
\label{sec:intro}

Electricity markets have become one of the central transmission mechanisms through which geopolitical shocks, fossil-fuel scarcity, renewable intermittency, and climate policy affect European economies. 
The European energy crisis following the Russian invasion of Ukraine showed that disruptions originating in fossil-fuel markets can rapidly propagate to wholesale electricity prices, with consequences for households, firms, inflation, public finances, and industrial competitiveness (\citealp{neri2025energyprices}; \citealp{acer2022market}). More recent tensions in the Middle East and the renewed concern about disruptions to oil and gas flows through the Strait of Hormuz confirm that energy-market fragility remains a major source of macroeconomic and policy risk. At the same time, as discussed by \cite{gianfreda2020comparing}, the expansion of wind and solar generation has changed the structure of European power markets.
Since electricity is only partially storable, renewables may lower prices when they are abundant, but their variability increases the importance of dispatchable generation when wind and solar output are low and residual demand must be met by fossil-fuel or other flexible technologies. 

These interactions matter for energy policy because they make the sources of electricity price movements difficult to identify from observed price comovements alone. For example, an upward co-movement of gas and electricity prices is not, by itself, evidence that an electricity-price innovation is a gas shock: it may instead reflect an oil- or coal-price shock that propagates through fuel-cost linkages and ultimately affects electricity prices. Similarly, it may be due to weak renewable supply, higher expected demand, or other disturbances. 

Disentangling these ``structural shocks'' is precisely {the objective and novelty} of our study. In particular, {we consider a structural vector autoregression (SVAR) to jointly model electricity market variables and} identify which structural shocks are mostly responsible for driving daily day-ahead electricity prices {in two different countries: Germany and Italy. The daily frequency of all the variables considered allows us to examine the transmission of different shocks over relatively short horizons, which is particularly relevant for day-ahead electricity markets}. We then investigate whether the transmission of renewable-supply, demand, fossil-fuel, carbon, and electricity-market-specific shocks {differs across countries and over regimes, and conduct a historical decomposition of spot electricity price forecast errors over the energy crisis and post-crisis periods.} {We exploit a novel identification strategy based on the merit-order ranking of generating units. This provides us with an economically plausible ordering of the shocks and motivates the required contemporaneous exclusion restrictions}.

Our analysis is motivated by the premise that different shocks would have different implications for market design and policy.  If price variation is mainly driven by fossil-fuel shocks, the relevant policy issue is not the merit-order mechanism itself, but the exposure of the electricity system to fossil-fuel price volatility. In this sense, energy security, fuel diversification, and long-term risk-allocation instruments---such as long-term contracts, Power Purchase Agreements (PPAs), and Contracts for Difference (CfDs)---address the root causes of volatility while preserving short-run market efficiency (see \citealp{acer2022market} for more details on the 2022 European gas price shock and \citealp{decarolis2026electricitymarketdesign} for the possible reforms of EU electricity market designs). 

If instead renewable-supply shocks account for a substantial fraction of price variation, then the relevant policy margin concerns flexibility, storage, transmission, and the management of intermittency. In systems with large shares of wind and solar generation, short-run prices remain important for dispatch, but additional flexibility is needed to absorb renewable production when it is abundant and to replace it when it is scarce \citep{joskow2019wholesale,newbery2018marketdesign,ambec2023electricityviews, decarolis2026electricitymarketdesign}. 
{In this regard, \citet{zerrahn2018storage} argue that the relevant issue is not to store all excess renewable production, but to combine storage, curtailment, and renewable capacity in an economically efficient way, helping translate abundant renewable generation into more persistent reductions in electricity-price exposure.} 

Finally, if electricity-market-specific shocks are significant, then regulatory interventions, bidding behavior, price exposure, and expectations may play a non-negligible role. The reason is that market rules and contractual arrangements affect how generators bid, how consumers hedge, how renewable producers are exposed to price risk, and how investors value flexible capacity \citep{fabra2023reforming,fabraImelda2023marketpower,schlecht2024financialcfd}.

{Following \cite{ravazzolo2025cap}, Germany and Italy provide a useful comparison; see Section 1 of the Supplementary Material for background information and IEA data on the domestic generation mix and on net electricity imports.}
Both countries were exposed to the European gas crisis and both participate in the integrated European electricity market, but their electricity systems differ in generation mix, renewable profiles, and price-setting mechanisms. Germany combines high renewable penetration, especially wind, with a historically important fossil-fuel component, particularly gas, coal, and lignite component. By contrast, despite a non-negligible contribution from domestic renewable generation, Italy remains heavily reliant on gas-fired power generation. This cross-country contrast also makes it possible to study whether common European fossil-fuel and carbon shocks transmit similarly across markets, or whether national electricity systems transform common shocks into different price dynamics.

To investigate these transmission channels, we estimate a structural vector autoregressive (SVAR) model using daily data from January 2021 to {September} 2025 for Germany and Italy. The system includes day-ahead electricity prices, one-day-ahead forecasts of load, wind generation, and solar generation, and daily prices of crude oil, natural gas, coal, and EU ETS carbon allowances. The identification strategy is recursive and is implemented by imposing short-run zero restrictions on the contemporaneous impact matrix of the VAR. The recursive ordering is motivated by the timing of day-ahead electricity market auctions and by the distinction between electricity-market variables, which are determined one day before physical delivery, and fossil-fuel and carbon prices, which are observed as daily closing prices on the delivery day. The model allows us to recover impulse responses, forecast error variance decompositions, and a crisis-specific historical decomposition of electricity prices in terms of the structural shocks.

We also conduct robustness checks using an alternative ordering of the variables in the model and replacing load forecasts with actual demand.
In particular, fossil fuel prices are ordered before  electricity prices and demand, reflecting the possibility that market operators submit their electricity-price bids on the basis of expectations about fossil fuel prices. Indeed, changing the ordering does not affect the main findings summarized below.

Our results can be summarized in three points. First, renewable-supply shocks lower electricity prices, consistently with the merit-order mechanism. This effect is especially visible for wind shocks in Germany, where wind explains a sizeable share of electricity-price variation \citep{sensfuss2008merit,hirth2013marketvalue,gasparella2023merit}. Second, fossil-fuel shocks become more important at longer horizons, with gas-price shocks playing the largest role. This is consistent with evidence that gas-fired generation often affects electricity prices more than its generation share alone would suggest \citep{zakeri2023role}. Third, the two countries differ substantially. In Germany, electricity-price variation is explained mainly by wind, gas-price, and electricity-market-specific shocks. In Italy, instead, gas-price and electricity-market-specific shocks dominate, while renewable-supply shocks play a smaller role.

The historical decomposition further illustrates that gas-price innovations explain a large share of the cumulative forecast errors in electricity prices during the European energy crisis; oil-price innovations also contribute, especially after the beginning of 2022. Around September 2022, the decomposition attributes part of the decline in electricity prices to negative electricity-market-specific shocks. This pattern is consistent with the announcement and anticipation of emergency EU interventions, including electricity-demand reduction targets, the inframarginal revenue cap, and discussions of a possible gas-price cap \citep{councilEU2022energymeeting,
councilEU2022emergencyintervention}\footnote{See \citet{councilEU2022emergencyintervention} for the
electricity-demand reduction targets and the cap on inframarginal electricity producers' revenues, and
\citet{councilEU2022gaspricecap} for the separate gas-market correction mechanism agreed in December 2022. For a scenario analysis on the effects of alternative gas-price caps, see \citet{ravazzolo2025cap}.}.

The remainder of the paper is organized as follows. In Section~\ref{sec:lit-rev}, we review the relevant literature on electricity markets and SVAR applications in energy markets. In Section~\ref{sec:data}, we describe the data and the construction of the daily variables. In Section~\ref{sec:svar}, we present the structural VAR and the identifying assumptions. In Section~\ref{sec:res}, we discuss the results of our empirical investigation, reporting the forecast error variance decompositions, impulse responses, and historical decompositions. We conclude in Section~\ref{sec:conclusion}.

\section{Literature review}
\label{sec:lit-rev}

This paper is related, first, to the structural VAR literature in macroeconomics and energy economics. Since \citet{sims1980} and \citet{blanchard1989}, SVARs have become a standard tool for policy analysis because they allow macroeconomic fluctuations to
be decomposed into shocks with different economic interpretations. This feature is particularly important in energy markets, where price movements may reflect different underlying disturbances. For example, \citet{hamilton2003} shows that the macroeconomic effects of oil-price movements depend on how the shock is measured, emphasizing nonlinearities and asymmetries between oil-price increases and decreases, while \citet{kilian2009} provides a decomposition of oil-price movements into
supply, aggregate-demand, and oil-specific demand shocks. \citet{baumeister2019} show that structural conclusions in the oil market depend on the identifying assumptions imposed to recover economically meaningful shocks. Recent evidence shows that the effect of energy-price shocks on inflation expectations is highly shock-specific. \citet{aastveit2023inflation} document that oil-price pass-through depends on the demand and supply forces driving the global oil market, while \citet{kilian2022oil} show that gasoline-price shocks can account for a substantial share of fluctuations in household inflation expectations. Related SVAR methods have also been applied to natural gas markets, for example by \citet{nick2014}. We follow the SVAR approach, but apply it to daily electricity markets, where renewable supply, expected demand, fossil-fuel prices, carbon prices, and electricity-market specific disturbances interact and contribute to the price-formation process.

A more proximate literature applies structural time-series methods directly to wholesale electricity markets. \citet{fezzi2010} estimate a structural asymmetric vector error-correction model for hourly day-ahead electricity markets and show that short-run demand is highly inelastic, although traded quantities respond to price levels and to supply-side disequilibria in an asymmetric way. \citet{bockers2013} estimate a structural VAR for the Spanish wholesale market to study how wind and solar generation affect the merit order and wholesale prices. \citet{ganepola2023electric} is particularly close to our identification problem, since the paper uses a short-run identified SVAR to distinguish fundamental shocks, associated with demand, supply, and generation costs, from electricity-price-specific shocks in the UK market; the results show that these electricity-specific shocks can account for a large fraction of price volatility. \citet{paschen2016} estimates a dynamic SVAR for the German day-ahead market and finds that positive wind and solar shocks reduce spot prices, with wind shocks displaying a more persistent negative effect than solar shocks. Finally, \citet{maciejowska2022} uses an SVAR for the German day-ahead and intraday markets in a portfolio management problem for a renewable-energy producer. These contributions show that structural time-series models are useful for studying electricity-price formation, but they usually focus on a single market, on renewable generation alone, or on trading and portfolio applications. Relative to this literature, we study the crisis and post-crisis periods, combine renewable forecasts with fossil-fuel and carbon prices, and compare the transmission of structural shocks in Germany and Italy.

The closest connections of this study are with the literature on the European energy crisis, the merit-order mechanism, and the role of gas in electricity-price formation. \citet{zakeri2023role} show that fossil-fuel technologies set European electricity prices much more often than their generation share would suggest, with
natural gas playing a particularly important price-setting role in several countries. \citet{gasparella2023merit} analyze merit-order and price-setting dynamics in European electricity markets, emphasizing that the role of gas depends on the
interaction between fuel prices, residual demand, and renewable generation. \citet{cevik2025shocked} study volatility spillovers across European electricity
markets and show that cross-border transmission became especially relevant in a period marked by Russia's invasion of Ukraine and the expansion of renewable energy.
Our paper is closest in spirit and data architecture to \citet{ravazzolo2025cap} and \citet{ravazzolo2025integration}, which study European electricity and energy-market dynamics during the recent crisis. Differently from these contributions, we provide a
daily, recursively identified SVAR decomposition centered on day-ahead electricity prices.

{Finally, exposure to energy-price shocks links electricity markets to the wider discussion on energy security and economic resilience. Dependence on fossil fuels can transmit fuel-price increases to electricity prices, with consequences extending beyond higher energy bills \citep{fabra2023reforming}. Indeed, \citet{neri2025energyprices} finds that energy-price increases during the 2021--22 crisis contributed to higher inflation and weaker economic activity in the euro area. In addition, \citet{anaya2025investment} discuss how squeezed profit margins, uncertainty and financing constraints can reduce investment, particularly in energy-intensive firms. Sharp movements in energy prices can also trigger margin calls on derivatives, requiring firms to provide additional cash or collateral and creating liquidity stress \citep{furtuna2022derivatives}.}

\section{Data description}
\label{sec:data}

Following \cite{ravazzolo2025cap}, we focus our analysis on Germany and Italy and source all series from the London Stock Exchange Group (LSEG) to construct a daily dataset that covers the period from January 2021 to {September} 2025. Specifically, we consider day-ahead electricity prices, forecasted load (i.e., electricity demand), forecasted wind and forecasted solar generation, and the prices of crude oil, gas, coal, and carbon emission allowances, where each series is transformed to logarithms.

We restrict the sample to weekdays, since fossil-fuel and carbon markets do not trade on weekends. Electricity-market variables are country-specific and are originally observed over lower-than-daily time periods, whereas fossil-fuel and carbon prices are common to both countries, as they correspond to contracts traded in international financial markets, and are observed at daily frequency. A comprehensive review of the key events related to the European energy crisis is available in \citet{ravazzolo2025integration}.

Day-ahead electricity prices are determined in the auction held on day $t-1$ for physical delivery on day $t$ and are expressed in euros per megawatt hour (EUR/MWh). For Germany, the electricity price is the day-ahead auction price for the German bidding zone, operated by EPEX SPOT. For Italy, the electricity price is the national day-ahead reference price. Specifically, over the pre-2025 part of the sample, this series corresponds to the PUN (\textit{Prezzo Unico Nazionale}); from 1 January 2025 onward, it corresponds to the PUN Index GME. Both series are national reference prices for electricity traded on the Italian Day-Ahead Market (\textit{Mercato del Giorno Prima}), although from 2025 the Italian market operates in an institutional setting in which zonal prices play a more direct role in settlement (\citealp{gme2024punindex}).

{Moreover, electricity prices differ across the two countries, as they can occasionally become negative in Germany when high renewable generation coincides with low demand. In contrast, Italian prices remain positive throughout our sample.
In particular, five of the six negative daily averages in Germany fall on weekends and are already excluded from our sample. To apply the log transformation, we treat the only remaining negative price, on 5 April 2021, as missing and exclude from the VAR any observation that uses it as a lag.}

For most of the sample, market time units are hourly. However, following the introduction of the 15-minute time unit in the Single Day-Ahead Coupling (beginning with delivery day 1 October 2025), prices are available at quarter-hourly frequency (\citealp{europeancommission2025quarterhourlymtu}). We aggregate sub-daily electricity prices to daily frequency by computing an arithmetic average over all time periods on the delivery day; the same aggregation principle is therefore applied before and after the transition from hourly to quarter-hourly prices.

Load, wind generation, and solar generation are one-day-ahead forecasted quantities. These forecasts come from the weather-information model of the European Centre for Medium-Range Weather Forecasts and, if unavailable, from the Global Forecast System (\citealp{ecmwfIFS}). Since these quantities are reported in megawatt hours, daily series are obtained by summing all available ``intraday'' quantities within each day. Daylight-saving transitions are handled mechanically by aggregating over the actual number of time periods available on each delivery day. Seasonal variation is particularly evident in the data and is therefore explicitly accounted for in the model specification.
{We use forecasted rather than realised demand
because the day-ahead price for delivery on day $t$ is set in the auction held on day $t-1$, when realised quantities for day $t$ are not yet observed. Market participants bid on the basis of the forecasts available before the auction closes, so these forecasts correspond to the information set relevant for day-ahead price formation. In particular, deviations of realised quantities from their forecasts materialise only after the auction and are settled in intraday and balancing markets.}

Fossil-fuel and carbon prices correspond to the front-month futures series for Brent crude oil, Dutch TTF natural gas, API2 coal delivered into the Amsterdam--Rotterdam--Antwerp region, and to the EU carbon allowances in the EU Emissions Trading System (EU ETS), included as daily closing prices. Gas and EU ETS allowance prices are already denominated in euros, whereas Brent crude oil and API2 coal prices are converted into euros using the daily EUR/USD closing exchange rate.
{
We measure fuel and carbon costs with daily market closing prices. Even when generators buy fuel under long-term contracts, hold inventories, or hedge their exposure, the marginal cost relevant to the day-ahead bid is the opportunity cost of the fuel, that is, the price at which it could be sold on the market. This is particularly true for gas-fired plants, whose dispatch decision is short-term (burning gas for power generation rather than selling it or using it for heating) and which generally do not hold storage capacity to accumulate gas over long horizons. The latest daily price therefore provides the most relevant measure of the fuel cost faced by generators.
}

\section{SVAR identification through the merit order}
\label{sec:svar}

We consider a daily vector autoregressive model of the German and Italian energy markets for $\mathbf{y}_t = (wind_t, solar_t, demand_t, electricity_t, oil_t, gas_t, coal_t, carbon_t)$, where all series are expressed in logs and sourced from LSEG. Specifically, $wind_t$ and $solar_t$ correspond to the 24-hour sum of one-step-ahead forecasts (that is, originating at time $t-1$) of wind and solar generation, $demand_t$ to the 24-hour sum of one-step-ahead forecasts of load, and $electricity_t$ is the 24-hour average of day-ahead electricity prices settled on day $t-1$ for delivery on day $t$. Conversely, $oil_t$, $gas_t$, $coal_t$, $carbon_t$ are daily ``close'' prices sampled at the end of day $t$. 
{We opt for a VAR in log-levels since it yields consistent estimates even in the presence of integrated and cointegrated variables, without requiring the cointegration rank to be specified (\citealp{sims1990inference}; \citealp{kilianlutkepohl2017}) and it is in line with the day-ahead electricity literature (e.g.,  \citealp{gianfreda2020comparing}; \citealp{weron2014electricity}).}
Thus, we consider a structural VAR representation with five lags separately for each country, that is
\begin{equation}
\label{section:svar:eq:svar}
    \mathbf{B}_0\mathbf{y}_t = \mathbf{C}\boldsymbol{\nu}_t + \sum_{i=1}^{5}\mathbf{B}_{i}\mathbf{y}_{t-i} + \mathbf{w}_t,
\end{equation}
where $\boldsymbol{\nu}_t$ is a vector of deterministic terms, including an intercept, seasonal harmonics, and dummies to account explicitly for calendar effects in the demand (i.e., national celebrations and holiday effects; see Appendix~\ref{sec:append_harmon} for a complete treatment of calendar effects), while $\mathbf{C}$ is the corresponding coefficient matrix. {To select the VAR lag order, we set the maximum lag order to 30 and choose five lags according to the Akaike information criterion (AIC). Moreover, for the impulse response analysis, we consider a robustness check where we increase the order to 10 lags (see Section S.4 in the Supplementary Material).}

$\mathbf{B}_{i}$ refers to the lag-specific coefficients matrix, while $\mathbf{w}_t$ denotes the vector of serially and mutually uncorrelated structural innovations:
\begin{equation}
\label{sec:svar:eq:w}
\mathbf{w}_t = 
\begin{pmatrix}
\begin{aligned}
&\mathbf{w}_t^{\textit{renewable supply shock}}      \\
&\mathbf{w}_t^{\textit{electricity market shock}}       \\
&\mathbf{w}_t^{\textit{fossil price shock}} \\
&w_t^{\textit{carbon price shock}}     \\
\end{aligned}
\end{pmatrix}
=
\begin{pmatrix}
\begin{array}{l l}
w_t^{\textit{wind supply shock}}      & \\[4pt]
w_t^{\textit{solar supply shock}}     & \\[4pt]
\hdashline
w_t^{\textit{electricity demand shock}}   & \\[4pt]
w_t^{\textit{electricity price shock}} & \\[4pt]
\hdashline
w_t^{\textit{oil price shock}}       & \\[4pt]
w_t^{\textit{gas price shock}}       & \\[4pt]
w_t^{\textit{coal price shock}}      & \\[4pt]
\hdashline
w_t^{\textit{carbon price shock}} &
\end{array}
\end{pmatrix}.
\end{equation}
Denote $\mathbf{u}_t$ as the reduced-form errors with covariance matrix $\Sigma_u = \mathbf{B}_0^{-1}\mathbf{B}_0^{-1\prime}$. From a statistical standpoint, we identify the parameters of $\mathbf{B}_0^{-1}$ by applying the Cholesky decomposition to $\Sigma_u$, thus imposing that $\mathbf{B}_0^{-1}$ is a lower triangular matrix with ones on the main diagonal. Then $\mathbf{u}_t$ can be recovered as $\mathbf{u}_t = \mathbf{B}_0^{-1}\mathbf{w}_t$. In economic terms, this identification strategy dictates a recursive causal ordering on the contemporaneous behavior of shocks.

\begin{figure}[!t]
\centering
\includegraphics[width=\textwidth]{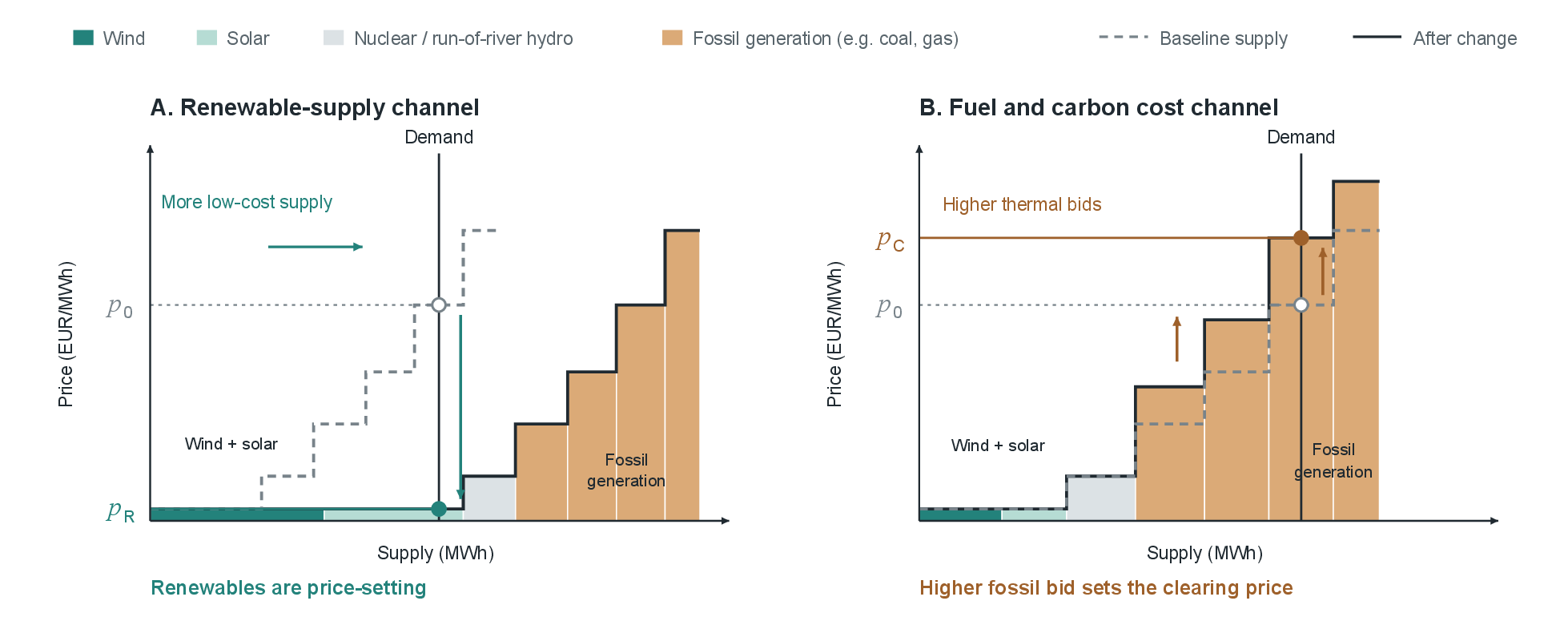}
\caption{Merit-order price formation with inelastic demand (see also \citealp{gasparella2023merit}). The dashed curve represents the initial supply curve, while the solid curve shows how greater renewable availability (Panel A) or higher fuel and carbon costs (Panel B) changes the marginal price at a given level of demand. Nuclear and hydro are included only for illustrative purposes.}
\label{fig:merit_order_identification}
\end{figure}

A key aspect of the identification strategy is the ordering of the variables, which is motivated as follows. We use forecasted variables as a useful way to distinguish two sets of variables according to the timing of information arrival within period $t$. The first block consists of $wind_t$, $solar_t$, $demand_t$, and $electricity_t$. These are either forecasts for delivery on day $t$ or the day-ahead electricity price settled on day $t-1$ for delivery on day $t$. The second block consists of $oil_t$, $gas_t$, $coal_t$, and $carbon_t$, which are instead observed as close prices at the end of day $t$. {Figure~\ref{fig:merit_order_identification} illustrates the price-setting mechanism and the timing of information that motivate our identifying assumptions.}

The ordering within the first and second blocks, as defined in Eq. \eqref{sec:svar:eq:w}, mainly reflects the information available when the day-ahead auction clears. Wind and solar generation are weather-driven, therefore renewable shocks can instantaneously affect electricity demand and prices, thus affecting market-clearing conditions via the merit-order criterion (\citealp{gasparella2023merit,acer2025key}). Notice that the exclusion restriction where shocks to solar generation do not affect wind generation should be read mainly as a requirement of the recursive structure and not as a peculiar causal ordering. We postulate that demand forecasts react to renewable forecasts and instantly affect electricity prices. This is consistent with the fact that, at the daily frequency, electricity demand is driven primarily by weather conditions, calendar effects, and expected economic activity, whereas market-clearing prices are determined by expected renewable generation, whose volume in relation to fossil fuels sets the marginal unit, and by electricity consumption. Therefore, we impose the mild economic restriction that increases (decreases) in electricity prices do not reduce (increase) demand immediately, but with at least a delay of a day.

This interpretation is consistent with recent European market episodes. Periods of abundant renewable generation and weak demand have been associated with frequent low and negative day-ahead prices, while periods of low wind and solar output have produced the opposite effect (\citealp{acer2025key}). A clear example is the German \textit{Dunkelflaute} episode in December 2024, when weak renewable generation, cold weather, and the need for thermal generation and imports were accompanied by large day-ahead price spikes (\citealp{acer2025key}). These episodes support the assumption that shocks to renewable generation and load can move electricity prices substantially, and that demand is inelastic in the short run to changes in market prices.

The shocks in the first and second blocks are predetermined with respect to the third and fourth blocks. The reason is the timing of market outcomes: oil, gas, coal, and carbon prices are measured as same-day close prices, whereas the day-ahead electricity price for delivery on day $t$ has already been settled on day $t-1$. {To illustrate this point, consider a generator expecting gas prices to rise: this expectation may already influence its bids for electricity delivery the following day. An unexpected change in gas prices after the auction has cleared, however, cannot alter the electricity price already determined.} The recursive ordering therefore imposes that same-day innovations in fossil and carbon markets do not feed back contemporaneously into the auction that has already cleared.

Within the third block, oil is the broadest and most liquid fossil-fuel market and is driven by global conditions rather than by local developments in the German and Italian electricity markets (\citealp{acer2022market, ecb2022eb4}). By contrast, gas and coal remain more regional markets due to transportation costs, infrastructure constraints, and the importance of regional benchmark contracts, even though gas markets have become more interconnected through LNG trade (\citealp{iea2025gas}). We assume that gas prices shocks affect coal prices contemporaneously, while ruling out contemporaneous effects in the opposite direction, and offer two related considerations. First, gas prices tend to transmit more directly to electricity prices because gas-fired units often set the marginal price in European wholesale markets (\citealp{gasparella2023merit,acer2022market}). Second, increases in gas prices create immediate incentives for gas-to-coal switching whenever technical and regulatory constraints allow it, so that coal prices are more plausibly downstream of gas through relative fuel-cost and substitution effects (\citealp{ec2022q2electricity,ecb2022eb4}). The European energy crisis provides a clear illustration of this mechanism: after Russia's invasion of Ukraine, oil, gas, and coal prices all rose sharply, gas became a key driver of wholesale electricity prices in Europe, and the increase in gas prices supported a temporary return to coal generation (\citealp{ecb2022eb4,ec2022q2electricity,acer2022market}). In our daily model, this does not mean that coal shocks cannot affect gas prices; rather, they are allowed to do so only with a delay of at least one day.

In the fourth block, we exclude that carbon shocks feed instantaneously into fossil prices. The rationale is that innovations in fuel prices can alter the relative profitability of gas- and coal-fired generation and therefore the demand for emission allowances within the EU ETS (\citealp{ecb2022eb3}). In particular, increases in gas prices may induce a shift toward coal, raising expected emissions and placing upward pressure on allowance prices (\citealp{ecb2022eb3,ec2022q2electricity}). This channel was visible during the European energy crisis, when high gas prices coincided with coal substitution and tighter carbon market conditions (\citealp{ec2022q2electricity,ecb2022eb3}). At the same time, the empirical link from coal and gas to carbon prices might not be unidirectional, so our model allows for effects in the opposite order after a day, e.g. via policy news, compliance demand, and broader macro-financial conditions (\citealp{ecb2022eb3}).

\section{Power play: the drivers of electricity-price shocks}
\label{sec:res}

We present the results for the German and Italian energy markets based on the merit-order specification introduced in Section~\ref{sec:svar}. Since our analysis is centered on electricity markets, we focus on the behavior of electricity prices in response to the identified structural shocks. {In the remainder of this section, we discuss the results along two complementary dimensions: differences between Germany and Italy, and changes between the crisis and post-crisis periods within each country.} {In Section S.5 of the Supplementary Material, we show that our findings remain essentially unchanged under an alternative ordering based on different assumptions. We also report results for the other variables in the VAR.}

\subsection{Forecast error variance decompositions}

Tables~\ref{tab:fevd_electricity_deu_full} and \ref{tab:fevd_electricity_ita_full} report the contribution of each of the eight shocks to the forecast error variance decomposition (FEVD) of electricity prices in Germany and Italy at different horizons\footnote{Our findings remain essentially unchanged when aggregating hourly variables over peak hours (8 a.m. to 8 p.m.) either using averages or sums. The corresponding results are available from the authors upon request.}. {First, we observe a common pattern across the two countries: electricity-market-specific shocks are more important at short horizons whereas fossil-fuel shocks gain relevance as the forecast horizon increases}. At the same time, the decomposition also reveals substantial cross-country differences. In Germany, most of the variation is attributed to wind supply shocks, gas price shocks, and electricity price shocks, which we interpret as broader electricity market specific shocks. In Italy, renewable-supply shocks play a much smaller role, while most of the variation is explained by gas price shocks and electricity price shocks.

\begin{table}[!t]
\centering
\small
\caption{Forecast error variance decomposition of electricity prices for Germany over the full sample.}
\label{tab:fevd_electricity_deu_full}
\begin{tabular*}{\textwidth}{@{\extracolsep{\fill}}lcccccccc@{}}
\toprule
Horizon & Wind & Solar & Demand & Electricity & Oil & Gas & Coal & CO$_2$ \\
\midrule
1 & 0.363 & 0.014 & 0.055 & 0.568 & 0.000 & 0.000 & 0.000 & 0.000 \\
5 & 0.354 & 0.024 & 0.065 & 0.481 & 0.011 & 0.056 & 0.004 & 0.005 \\
10 & 0.322 & 0.024 & 0.068 & 0.438 & 0.015 & 0.121 & 0.005 & 0.007 \\
20 & 0.288 & 0.027 & 0.065 & 0.393 & 0.024 & 0.187 & 0.007 & 0.010 \\
30 & 0.267 & 0.028 & 0.061 & 0.365 & 0.033 & 0.224 & 0.011 & 0.011 \\
\bottomrule
\end{tabular*}
\end{table}

\begin{table}[!t]
\centering
\small
\caption{Forecast error variance decomposition of electricity prices for Italy over the full sample.}
\label{tab:fevd_electricity_ita_full}
\begin{tabular*}{\textwidth}{@{\extracolsep{\fill}}lcccccccc@{}}
\toprule
Horizon & Wind & Solar & Demand & Electricity & Oil & Gas & Coal & CO$_2$ \\
\midrule
1 & 0.063 & 0.041 & 0.070 & 0.825 & 0.000 & 0.000 & 0.000 & 0.000 \\
5 & 0.043 & 0.022 & 0.062 & 0.536 & 0.005 & 0.317 & 0.008 & 0.007 \\
10 & 0.031 & 0.017 & 0.049 & 0.397 & 0.011 & 0.472 & 0.014 & 0.009 \\
20 & 0.024 & 0.016 & 0.036 & 0.278 & 0.028 & 0.581 & 0.023 & 0.014 \\
30 & 0.021 & 0.016 & 0.031 & 0.226 & 0.041 & 0.614 & 0.032 & 0.020 \\
\bottomrule
\end{tabular*}
\end{table}


These results are consistent with the central role of the merit-order mechanism in electricity-price formation, especially when fossil-fuel technologies represent the marginal units, and in particular with the conclusions of \citet{zakeri2023role}. They estimate that in 2021, before Russia's full-scale invasion of Ukraine, gas-fired generation set day-ahead electricity prices for about 80\% of hours in Italy, whereas in Germany coal remained the dominant marginal fossil technology, setting prices for approximately 48\% of hours, compared with about 24\% for gas. 

Then we split the sample and consider the energy-crisis period, {from September 2021 to February 2023}, and a post-crisis period, {from March 2023 to September 2025}. Figure S.1 in the Supplementary Material clearly shows the sharp increase in day-ahead electricity prices during the energy crisis, followed by a return to relatively lower price levels in the post-crisis period.

\begin{table}[!t]
\centering
\small
\caption{Forecast error variance decomposition of electricity prices for Germany over the energy-crisis (September 2021--February 2023) and post-crisis sample (March 2023--September 2025). The table reports the bootstrap-mean contribution of selected shocks.}
\label{tab:fevd_electricity_deu_split}
\begin{tabular*}{\textwidth}{@{\extracolsep{\fill}}lcccccc@{}}
\toprule
& \multicolumn{3}{c}{Energy crisis} & \multicolumn{3}{c}{Post-crisis} \\
\cmidrule(lr){2-4}\cmidrule(lr){5-7}
Horizon & Wind & Electricity & Gas & Wind & Electricity & Gas \\
\midrule
1 & 0.442 & 0.456 & 0.000 & 0.378 & 0.566 & 0.000 \\
5 & 0.415 & 0.335 & 0.071 & 0.380 & 0.495 & 0.012 \\
10 & 0.366 & 0.306 & 0.120 & 0.360 & 0.472 & 0.037 \\
20 & 0.350 & 0.290 & 0.121 & 0.348 & 0.456 & 0.055 \\
30 & 0.347 & 0.286 & 0.121 & 0.344 & 0.450 & 0.060 \\
\bottomrule
\end{tabular*}
\end{table}

\begin{table}[!t]
\centering
\small
\caption{Forecast error variance decomposition of electricity prices for Italy over the energy-crisis (September 2021--February 2023) and post-crisis sample (March 2023--September 2025). The table reports the bootstrap-mean contribution of selected shocks.}
\label{tab:fevd_electricity_ita_split}
\begin{tabular*}{\textwidth}{@{\extracolsep{\fill}}lcccccc@{}}
\toprule
& \multicolumn{3}{c}{Energy crisis} & \multicolumn{3}{c}{Post-crisis} \\
\cmidrule(lr){2-4}\cmidrule(lr){5-7}
Horizon & Wind & Electricity & Gas & Wind & Electricity & Gas \\
\midrule
1 & 0.007 & 0.928 & 0.000 & 0.145 & 0.741 & 0.000 \\
5 & 0.022 & 0.529 & 0.358 & 0.147 & 0.534 & 0.167 \\
10 & 0.095 & 0.394 & 0.366 & 0.119 & 0.389 & 0.319 \\
20 & 0.106 & 0.343 & 0.344 & 0.100 & 0.301 & 0.399 \\
30 & 0.104 & 0.334 & 0.337 & 0.094 & 0.277 & 0.409 \\
\bottomrule
\end{tabular*}
\end{table}


Table~\ref{tab:fevd_electricity_deu_split} reports the FEVD for Germany separately over these two sub-samples. The comparison reveals differences between the crisis and post-crisis periods in the contributions of wind, electricity-market-specific, and gas price shocks. During the crisis, wind explains the largest share of the variance of electricity prices at short and medium horizons. Daily uncertainty in electricity prices is therefore more strongly associated with wind availability and electricity-market-specific shocks than with other innovations. {In the post-crisis period, after electricity prices reverted to lower levels}, the variance decomposition shifts toward electricity-market-specific shocks, {while the contribution of gas-price shocks at longer horizons is roughly half its crisis-period share. Wind nevertheless continues to account for approximately one-third or more of electricity-price variability across all horizons}. 

{The smaller contribution of gas may partly reflect the easing of electricity-supply constraints after the crisis. In 2023, according to \cite{bnetza2024marketdata}, Germany recorded higher wind generation, lower demand, and increased commercial electricity imports, while \cite{acer2025key} documents that the recovery of French nuclear output also improved regional supply conditions. Nonetheless, wind availability continues to determine how much demand must be met by other sources: when wind output falls, prices depend on the availability and cost of the generation and imports needed to replace it. Therefore, unexpected changes in plant availability or cross-border supply may be captured as electricity-market-specific shocks. The German electricity price spikes of December 2024 provide a clear illustration: low wind generation coincided with higher demand during cold weather and substantial outages at fossil-fuel power plants \citep{acer2025key}.}

Moving to Italy, Table~\ref{tab:fevd_electricity_ita_split} points to a different pattern, with renewables playing a smaller role and gas becoming more central, particularly during the crisis period. In the post-crisis period, the FEVD remains strongly gas-intensive {especially at horizons of 20 and 30}, indicating that the end of the energy crisis did not decouple Italian electricity prices from gas.

{GME's annual reports help interpret these results in more detail. The smaller gas contributions at horizons 5 and 10 partly reflect the recovery of hydropower and increased electricity imports in 2023, which provided alternatives to thermal generation \citep[p.~25]{gme2023report}. However, as discussed above, producing less electricity from gas does not necessarily diminish its role in setting prices: electricity sales by combined-cycle plants declined from 96.8 TWh in 2023 to 93.8 TWh in 2024, while their volume-weighted marginal-technology index increased from 60.2\% to 61.4\% \citep[p.~20]{gme2024report}. Moreover, constraints on imports can prevent cheaper foreign electricity from displacing domestic marginal generation, while the retreat of coal may also reduce competition from another thermal source.}

\subsection{Impulse response analysis}
\begin{figure}[!t]
\centering

\begin{subfigure}[b]{0.47\textwidth}
\centering
\includegraphics[width=\linewidth]{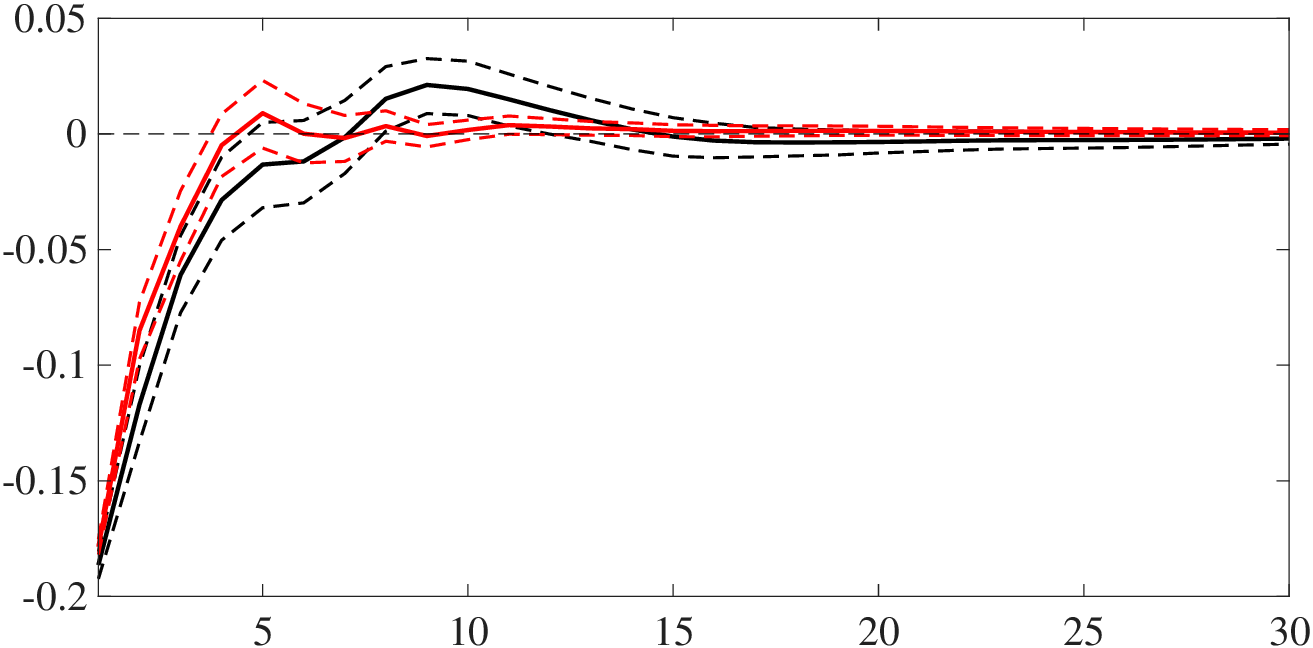}
\caption{Germany: wind}
\label{fig:period:deu:wind}
\end{subfigure}
\hfill
\begin{subfigure}[b]{0.47\textwidth}
\centering
\includegraphics[width=\linewidth]{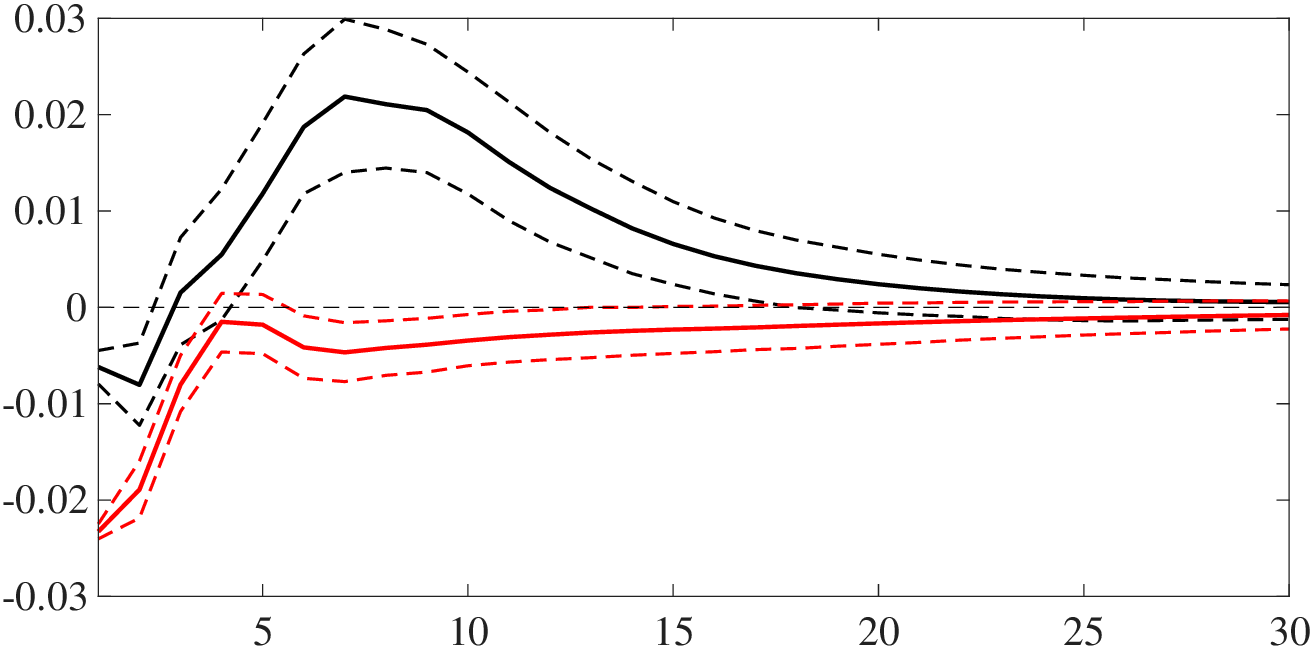}
\caption{Italy: wind}
\label{fig:period:ita:wind}
\end{subfigure}

\medskip

\begin{subfigure}[b]{0.47\textwidth}
\centering
\includegraphics[width=\linewidth]{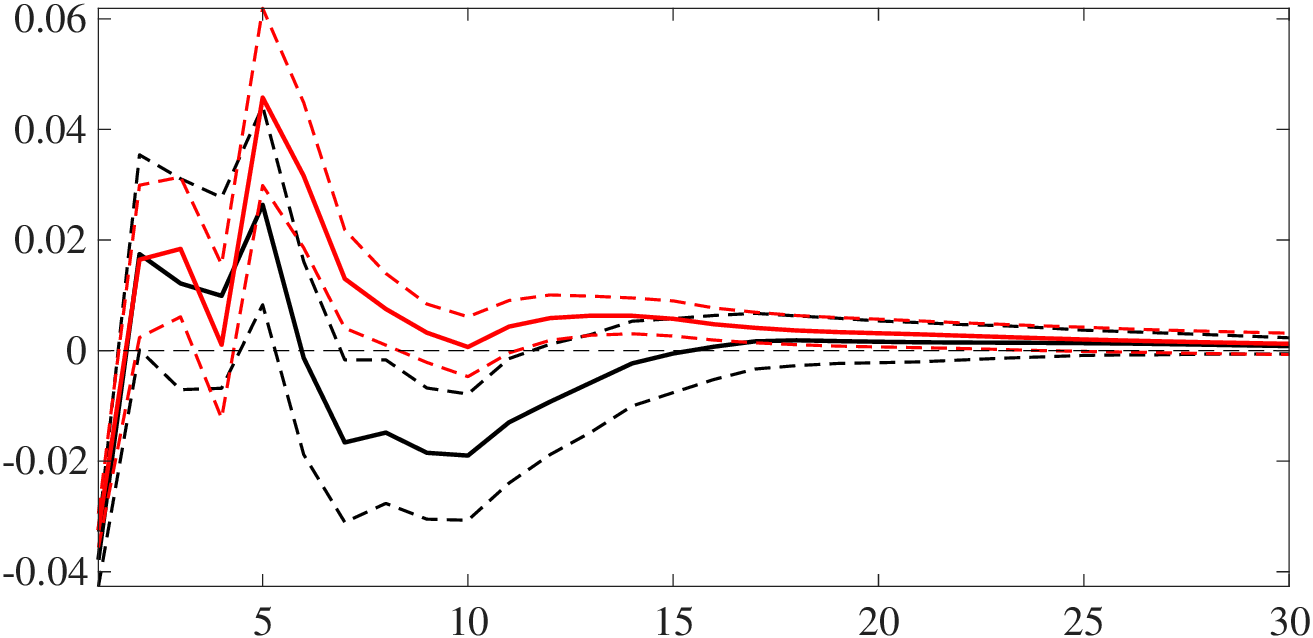}
\caption{Germany: solar}
\label{fig:period:deu:solar}
\end{subfigure}
\hfill
\begin{subfigure}[b]{0.47\textwidth}
\centering
\includegraphics[width=\linewidth]{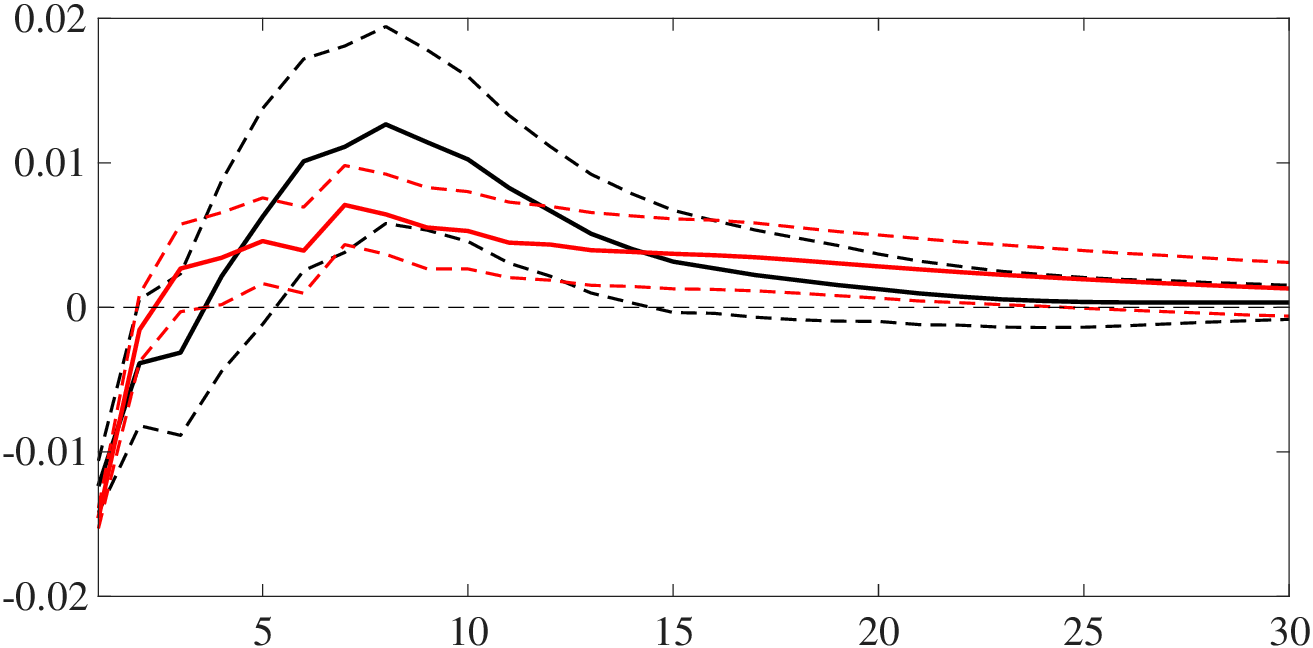}
\caption{Italy: solar}
\label{fig:period:ita:solar}
\end{subfigure}

\medskip

\begin{subfigure}[b]{0.47\textwidth}
\centering
\includegraphics[width=\linewidth]{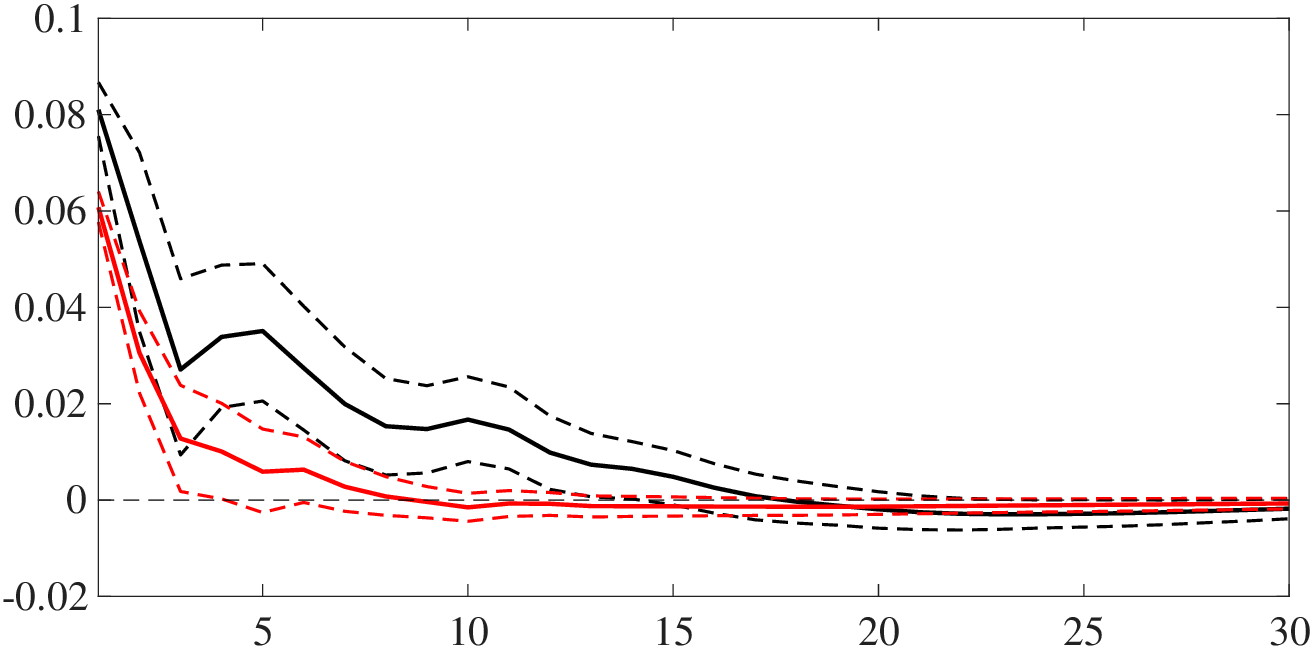}
\caption{Germany: demand}
\label{fig:period:deu:demand}
\end{subfigure}
\hfill
\begin{subfigure}[b]{0.47\textwidth}
\centering
\includegraphics[width=\linewidth]{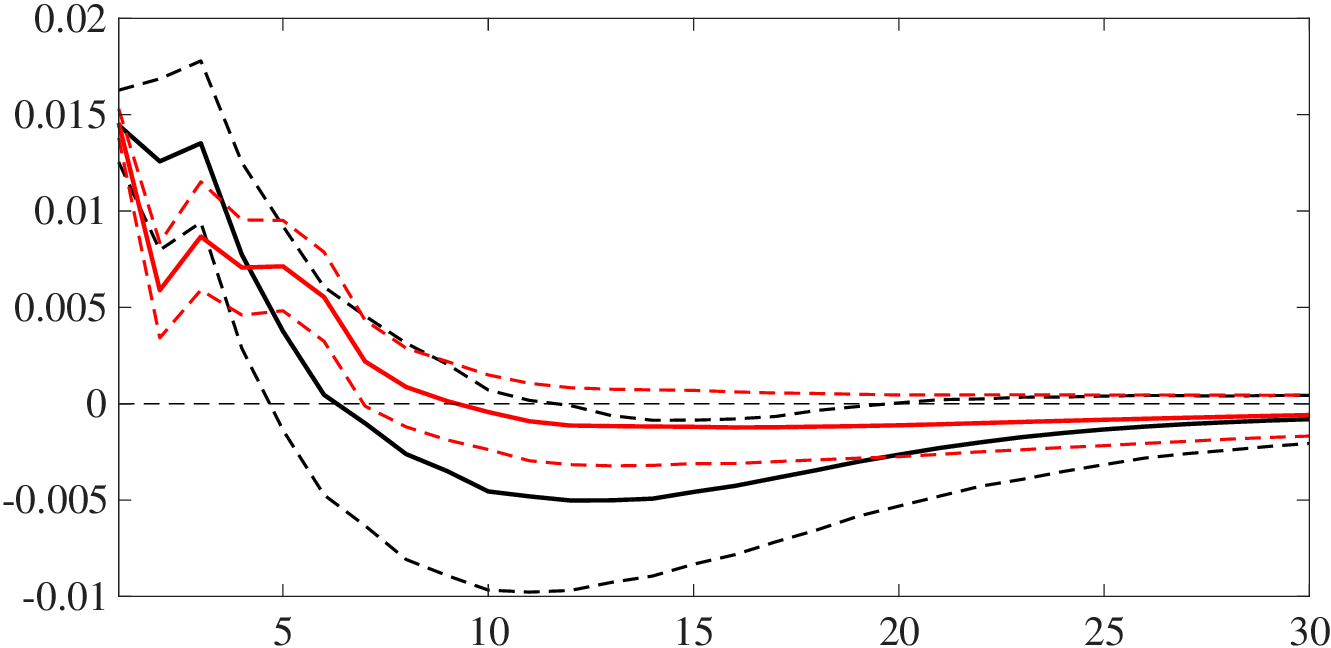}
\caption{Italy: demand}
\label{fig:period:ita:demand}
\end{subfigure}

\medskip

\begin{subfigure}[b]{0.47\textwidth}
\centering
\includegraphics[width=\linewidth]{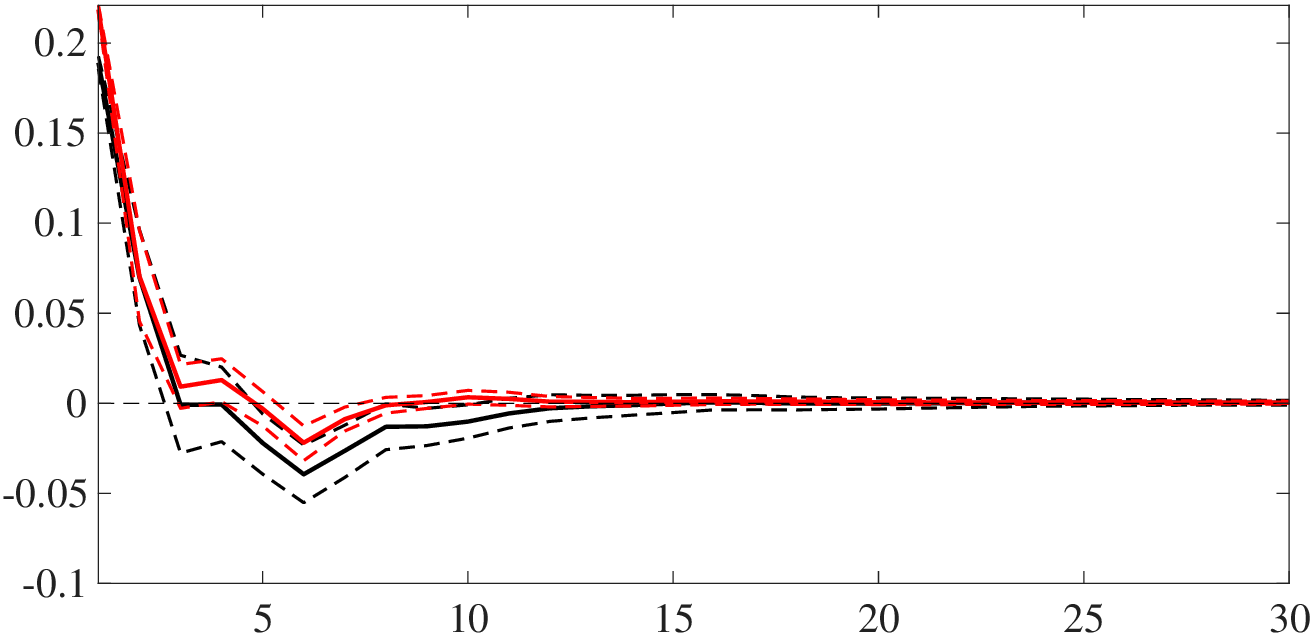}
\caption{Germany: electricity}
\label{fig:period:deu:electricity}
\end{subfigure}
\hfill
\begin{subfigure}[b]{0.47\textwidth}
\centering
\includegraphics[width=\linewidth]{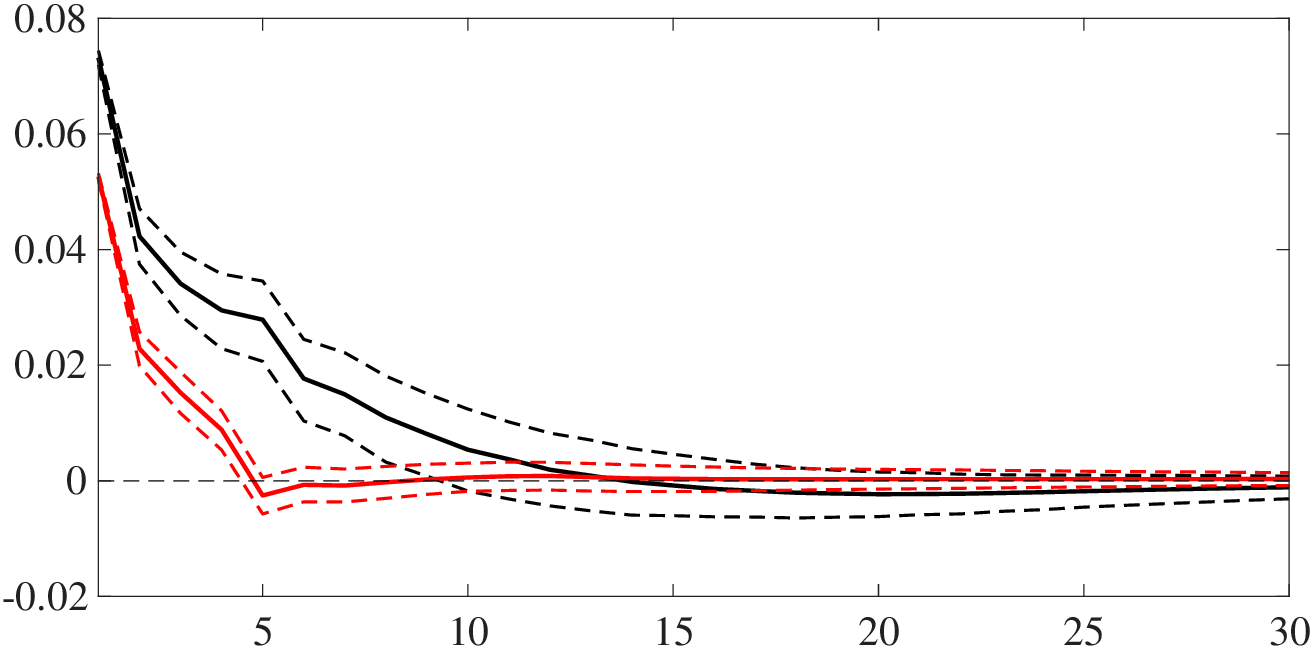}
\caption{Italy: electricity}
\label{fig:period:ita:electricity}
\end{subfigure}

\caption{{Impulse response analysis of electricity prices to structural shocks during the crisis (black) and post-crisis (red) periods over horizons up to 30 days. The solid lines correspond to the response to one-standard-deviation shocks, while the dotted lines represent 68\% block-wild-bootstrap intervals. }}
\label{fig:period:non-fossils}

\end{figure}

\begin{figure}[!t]
\centering

\begin{subfigure}[b]{0.47\textwidth}
\centering
\includegraphics[width=\linewidth]{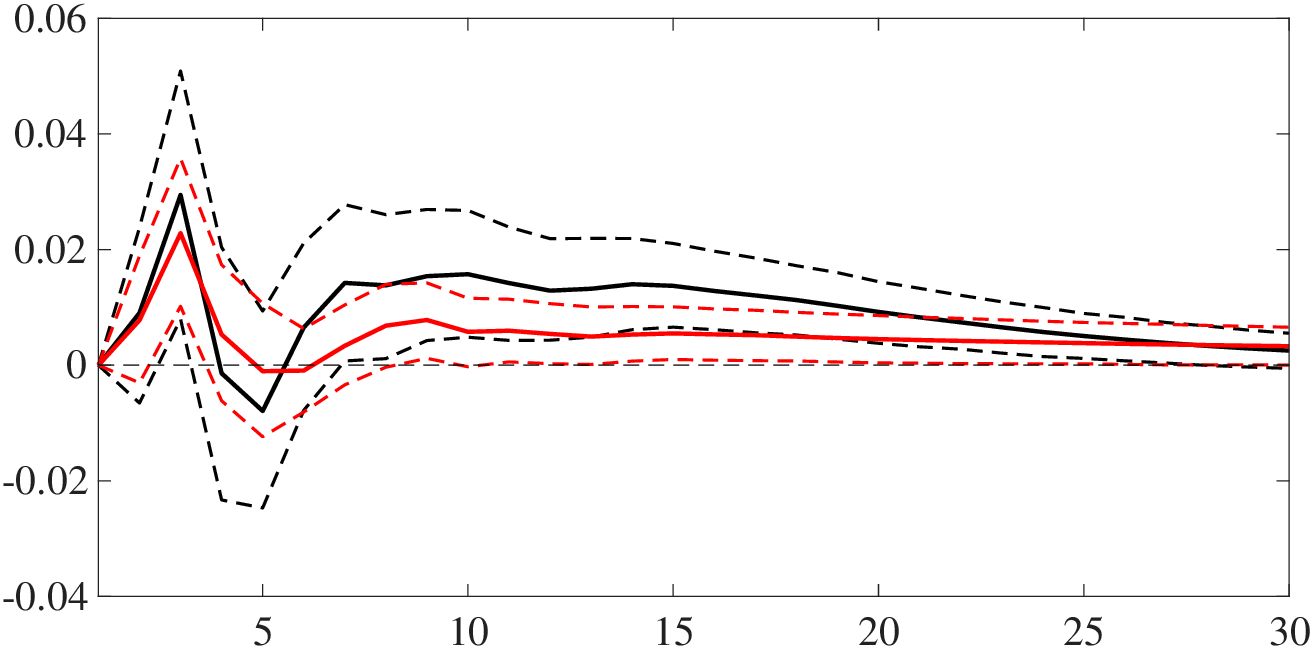}
\caption{Germany: oil}
\label{fig:period:deu:oil}
\end{subfigure}
\hfill
\begin{subfigure}[b]{0.47\textwidth}
\centering
\includegraphics[width=\linewidth]{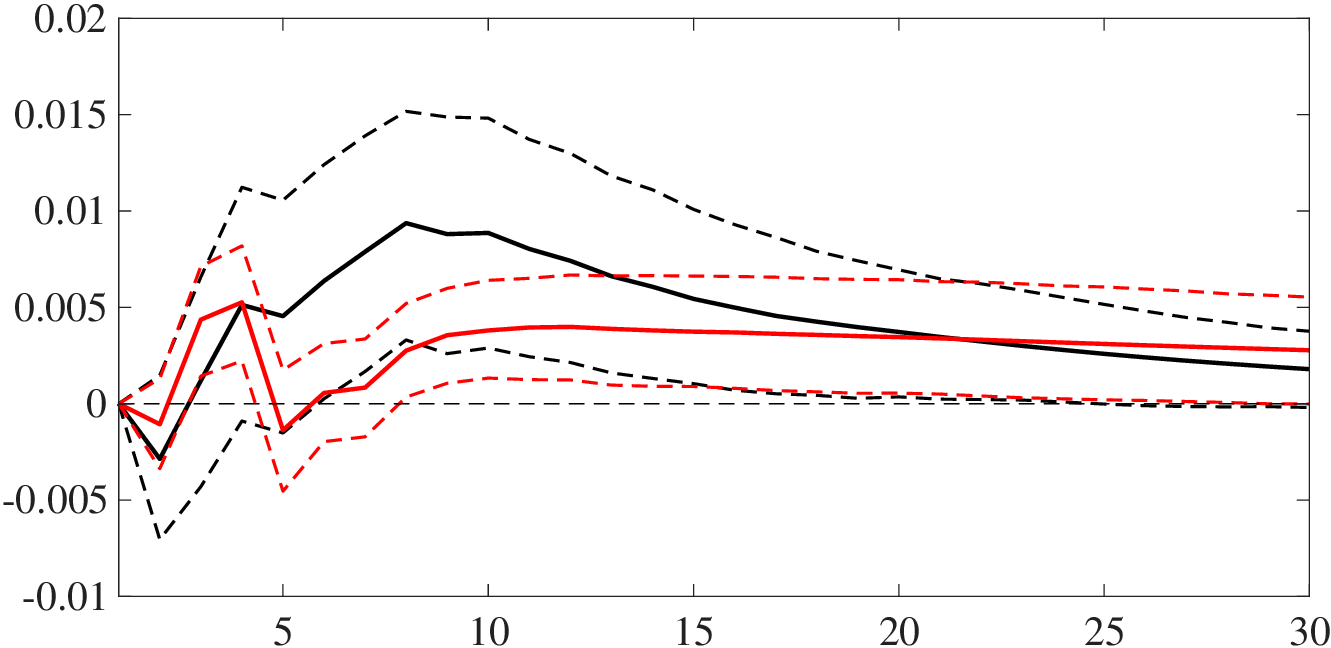}
\caption{Italy: oil}
\label{fig:period:ita:oil}
\end{subfigure}

\medskip

\begin{subfigure}[b]{0.47\textwidth}
\centering
\includegraphics[width=\linewidth]{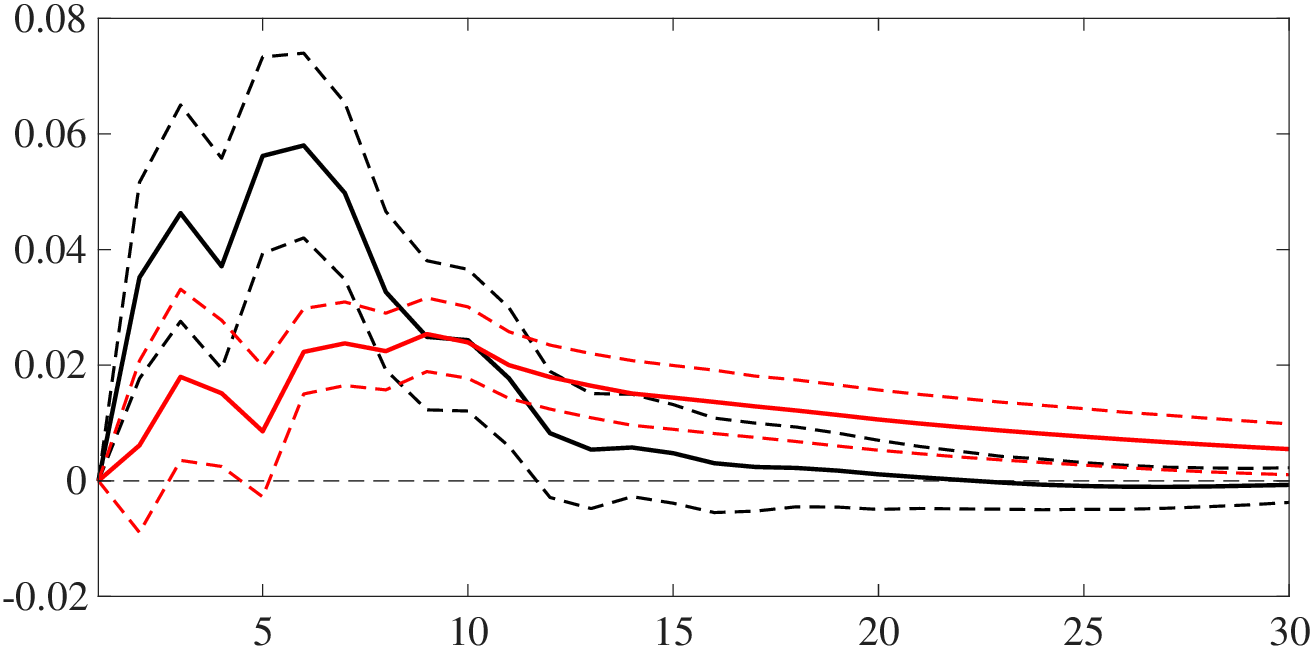}
\caption{Germany: gas}
\label{fig:period:deu:gas}
\end{subfigure}
\hfill
\begin{subfigure}[b]{0.47\textwidth}
\centering
\includegraphics[width=\linewidth]{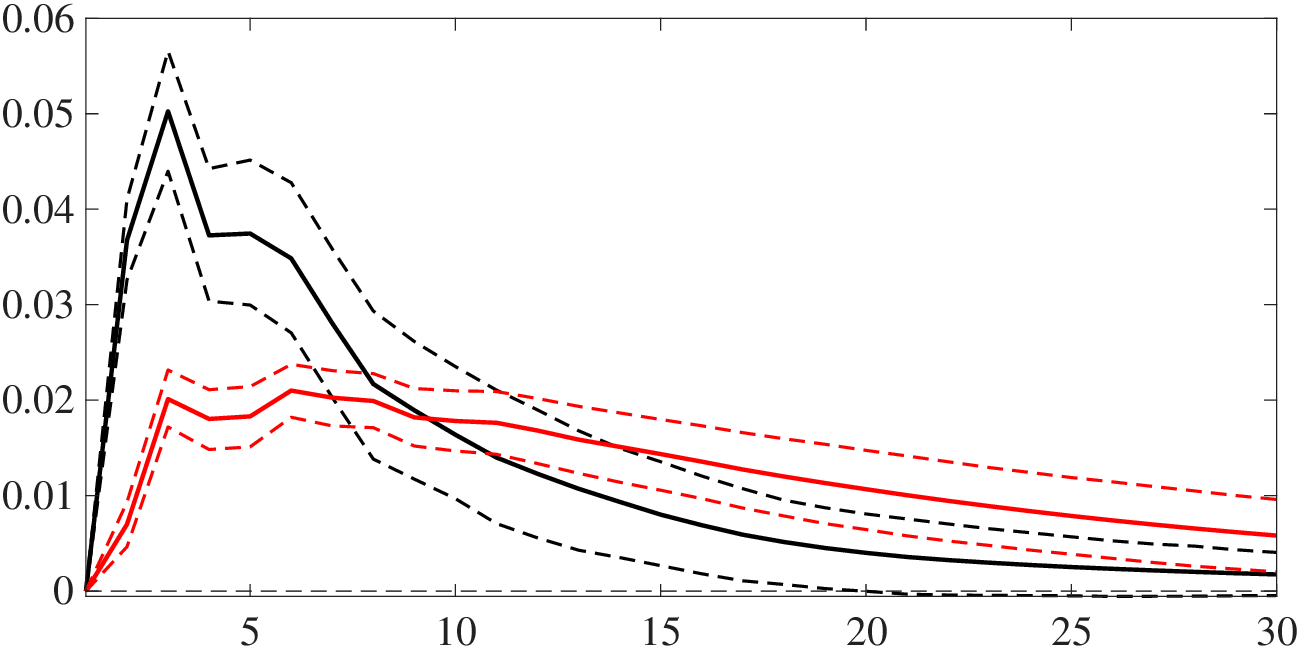}
\caption{Italy: gas}
\label{fig:period:ita:gas}
\end{subfigure}

\medskip

\begin{subfigure}[b]{0.47\textwidth}
\centering
\includegraphics[width=\linewidth]{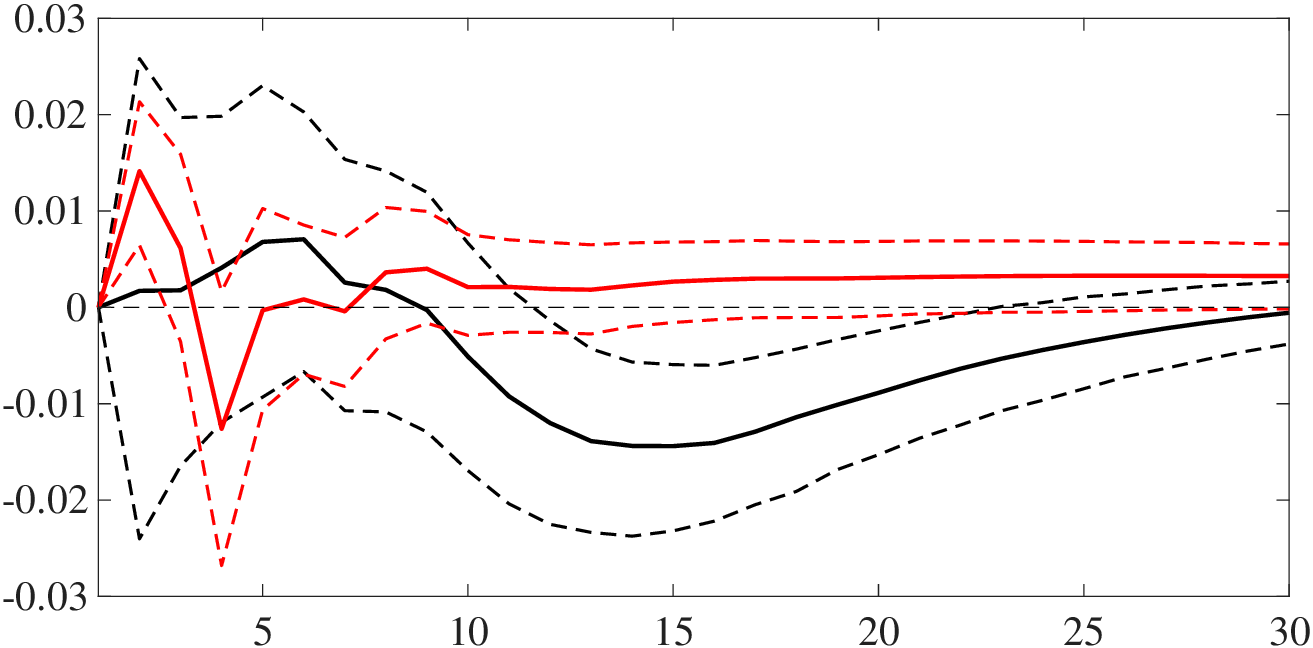}
\caption{Germany: coal}
\label{fig:period:deu:coal}
\end{subfigure}
\hfill
\begin{subfigure}[b]{0.47\textwidth}
\centering
\includegraphics[width=\linewidth]{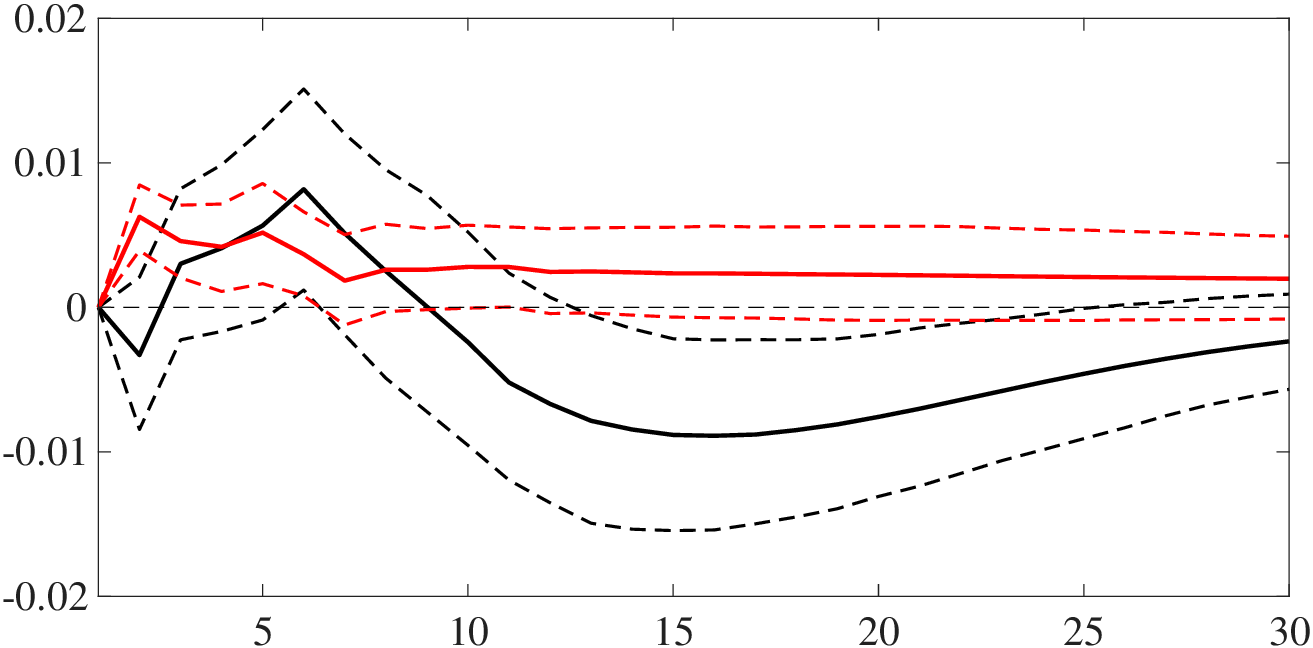}
\caption{Italy: coal}
\label{fig:period:ita:coal}
\end{subfigure}

\medskip

\begin{subfigure}[b]{0.47\textwidth}
\centering
\includegraphics[width=\linewidth]{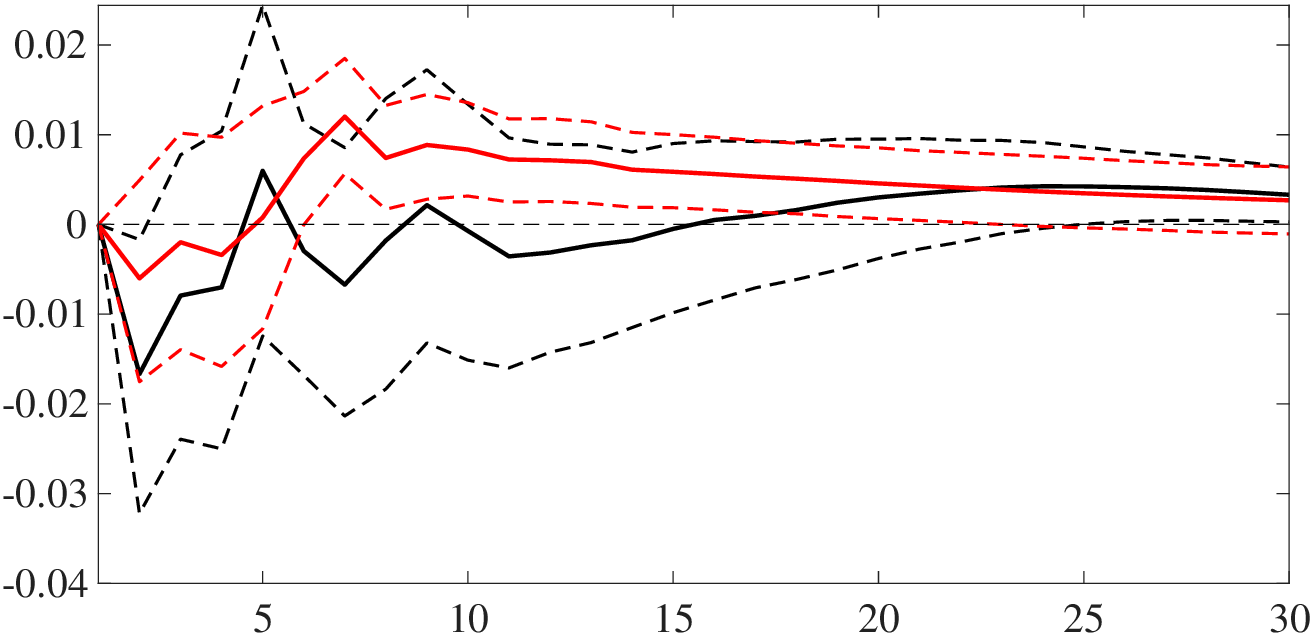}
\caption{Germany: carbon}
\label{fig:period:deu:co2}
\end{subfigure}
\hfill
\begin{subfigure}[b]{0.47\textwidth}
\centering
\includegraphics[width=\linewidth]{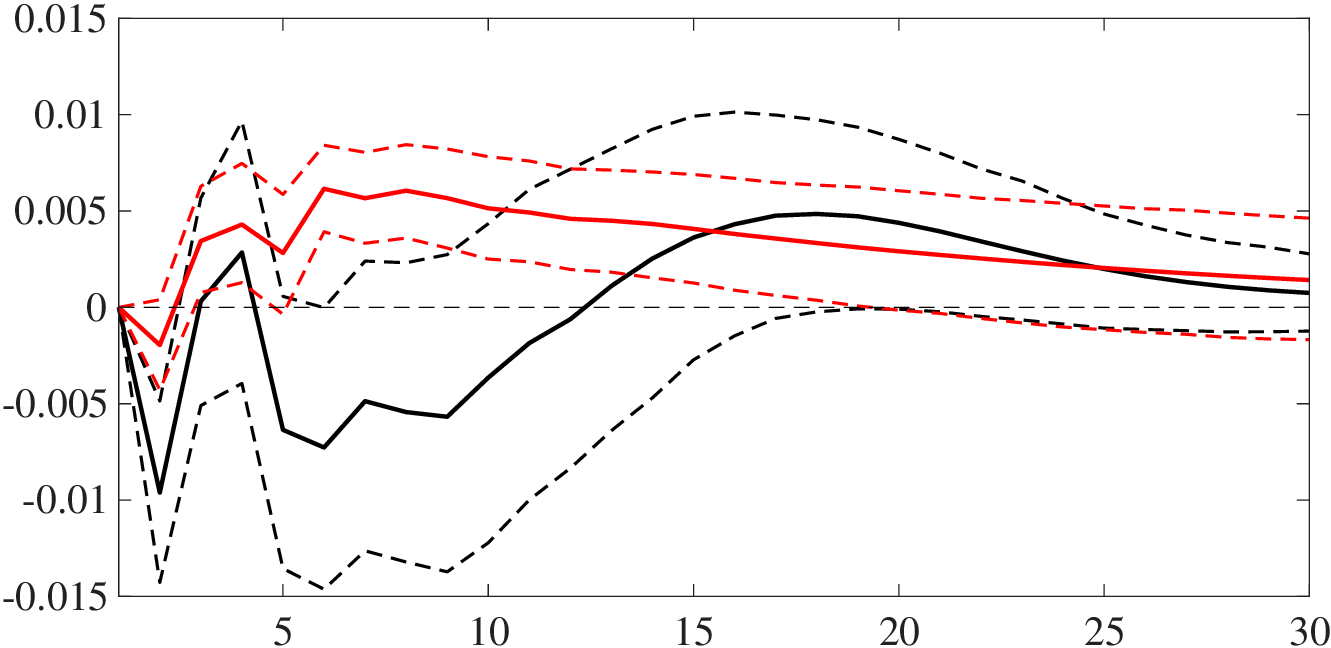}
\caption{Italy: carbon}
\label{fig:period:ita:co2}
\end{subfigure}

\caption[]{Impulse response analysis (continued).}
\label{fig:period:fossils}

\end{figure}

The impulse responses of electricity prices to one-standard-deviation structural shocks reported in Figures~\ref{fig:period:non-fossils} and \ref{fig:period:fossils} {allow us to assess how these responses differ between the crisis and post-crisis periods}. Since all endogenous variables are expressed in logs, the responses can be interpreted approximately as percentage changes in the corresponding variables in levels; {notice also that the shocks are scaled by their country- and period-specific standard deviations}. Moreover, since the variables in the VAR display substantial persistence, the impulse responses are not required to vanish quickly. We report responses up to 30 days ahead, a horizon where most responses have either stabilized or substantially flattened.

The estimated responses of day-ahead electricity prices display some common patterns across Germany and Italy, but also reveal several important differences (Panels~\ref{fig:period:deu:wind}--\ref{fig:period:ita:wind}). On impact, forecasted wind-supply shocks reduce German electricity prices {strongly both during and after the crisis ($>10\%$)}. {For Italy, this effect differs across periods: during the crisis, wind-supply shocks increase electricity prices in the short run before gradually returning to zero, whereas the post-crisis response is negative.} {The response to forecasted solar-supply shocks is initially negative in both countries and periods, while it becomes positive at intermediate horizons (Panels~\ref{fig:period:deu:solar}--\ref{fig:period:ita:solar}). Interestingly, in Germany the crisis response turns negative again approximately between horizons 5 and 10.}

{The \citet{bnetza2021minimum} documents that the strong initial wind response may be amplified by conventional generation that cannot readily adjust, for instance because of heat-supply obligations or to help maintain grid stability. However, a favourable wind forecast is not a permanent increase in generating capacity as its price effect depends on how long the additional wind persists; as discussed by \cite{paschen2016}, its duration also depends on the persistence of the underlying generation conditions. On the other hand, the subsequent rebounds are less straightforward. \cite{bockers2013} show that one possible mechanism is a change in the composition of conventional generation, with reduced operation of mid-merit plants but a continuing need for more flexible and expensive units.}

{The responses to forecasted electricity-demand shocks are positive in the short run and, in general, more comparable across the two periods (Panels~\ref{fig:period:deu:demand}--\ref{fig:period:ita:demand}). In contrast to Germany, the Italian responses become negative at intermediate horizons, and are statistically significant at some of these horizons; at longer horizons, in contrast, the responses are not statistically different from zero. On one hand, the positive initial response reflects movement towards more expensive supply, while its subsequent decline can arise because the unexpected demand increase subsides, additional supply becomes available, or both.}

{To conclude the set of predetermined variables, the responses to electricity-market shocks are positive on impact in both countries and then vanish quickly towards zero at relatively short horizons; their effect lasts somewhat longer in Italy during the crisis (Panels~\ref{fig:period:deu:electricity}--\ref{fig:period:ita:electricity}). Similarly to demand, the decay of electricity-market-specific shocks is rapid, compatibly with temporary changes in available capacity or bidding conditions. In Italy, for example, \cite{ternaAdequacyComparison} documents such circumstances in July 2022, when heat-related restrictions and outages reduced thermal availability, while hydropower and supplies from neighbouring countries were also constrained. Under these conditions, the market may have fewer alternatives with which to absorb a temporary shock.}

The remaining responses are subject to the short-run zero restrictions imposed in the identification scheme: fossil-fuel and carbon price shocks do not affect day-ahead electricity prices contemporaneously on the electricity delivery day. Consequently, their impulse responses are zero by construction on impact. {An important finding is that the responses in the split-sample analysis vanish more quickly, yet not necessarily approaching zero, but remain more persistent in the full-sample analysis.}

A clear pattern emerges: the responses to oil and gas price shocks are {predominantly positive after impact, with larger peaks during the crisis} (Panels~\ref{fig:period:deu:oil}--\ref{fig:period:ita:oil} and \ref{fig:period:deu:gas}--\ref{fig:period:ita:gas}, respectively). In particular, the {crisis} response of electricity prices to gas price shocks peaks after about {one week in Germany and three days in Italy}. {The post-crisis peaks are smaller, but the responses decline more slowly. In Italy, the post-crisis response exceeds its crisis counterpart from around horizon 10 onward.} By contrast, the effect of oil price shocks {is generally smaller and declines slowly after its early fluctuations, but remains persistent.}

{As discussed earlier, the pass-through of gas-price changes to wholesale electricity prices is well documented \citep{fezzi2009structural, guo2021cost, ravazzolo2025cap}. Interestingly, in our results their effects persist for longer after the crisis, a pattern  consistent with higher-cost units more often setting the marginal price. For oil, there exists an additional transmission channel through substitution between oil and gas, outside electricity generation: \cite{iea2023gasdemand} documents substantial gas-to-oil switching in European industry during 2022, suggesting that higher oil prices could weaken this alternative to gas use, thereby supporting gas demand and indirectly electricity prices.}

{At first glance, our finding of little evidence of a sustained electricity price shocks (Panels~\ref{fig:period:deu:coal}--\ref{fig:period:ita:coal}) may seem inconsistent with the substitution mechanisms discussed in in Section~\ref{sec:svar}.}
During the crisis, their effect is negative only at intermediate horizons and statistically different from zero in both countries. In the post-crisis period, the intervals generally include zero beyond the short run.

{However, a weak response of spot electricity prices does not necessarily contradict substitution between coal and gas in electricity generation. Indeed, during parts of the crisis, gas-fired generation was substantially more expensive than coal-fired generation \citep{ec2022q2electricity}. Under these conditions, an increase in coal costs could leave coal-fired plants cheaper than gas-fired plants, with little effect on electricity prices when gas remained the price-setting technology.}

{To conclude, we consider the responses to carbon price shocks} (Panels~\ref{fig:period:deu:co2}--\ref{fig:period:ita:co2}). {During the crisis, in both countries they are initially negative; in Germany, they quickly approach zero while the Italian responses remain negative for several observations before turning positive shortly after horizon 10. After the crisis, both responses are predominantly positive after the first few horizons and then gradually decline.}

{\citet{esma2022} offers useful context for interpreting the less straightforward difference between the IRFs in the crisis and post-crisis periods. In particular, one possible explanation is that, as gas prices surged, market participants may have sold emission allowances to meet immediate cash payments required on their energy derivatives, thus depressing carbon prices even while fuel costs were pushing electricity prices upwards. For example, \cite{furtuna2022derivatives} note how German energy firms faced substantial liquidity needs arising from margin calls. However, the crisis-period confidence intervals are wide and generally include zero, limiting the conclusions that can be drawn.}

\subsection{Historical decomposition}
\begin{figure}[!t]
    \centering

    \begin{subfigure}[b]{0.48\textwidth}
        \centering
        \includegraphics[width=\linewidth]{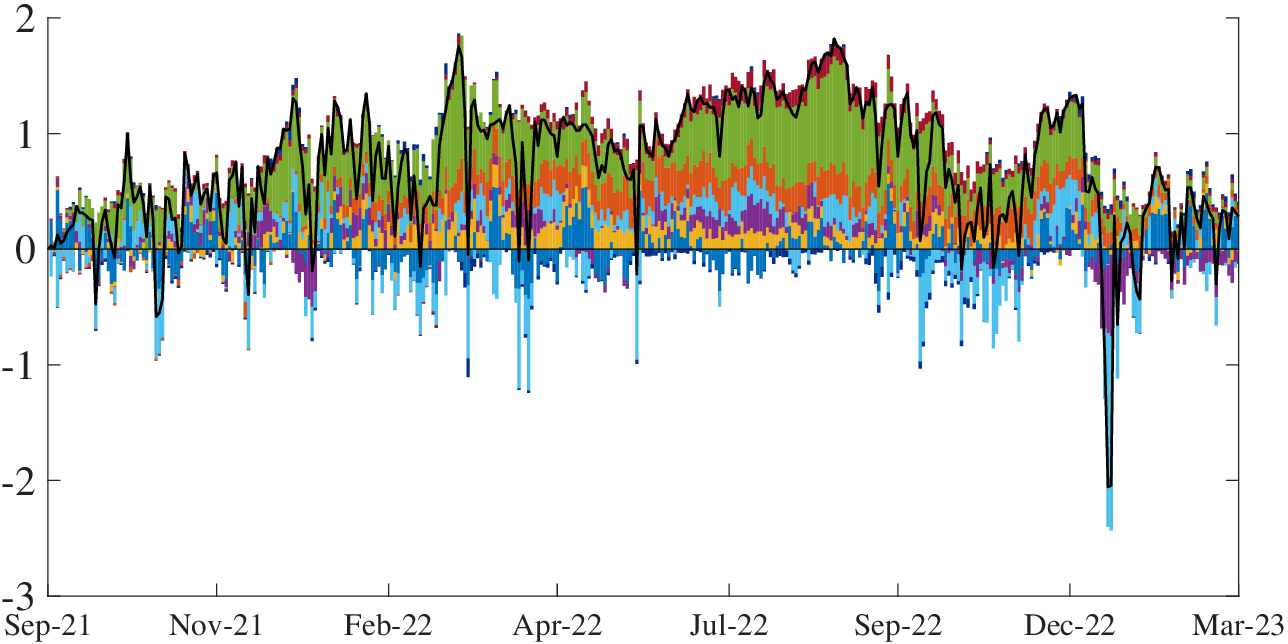}
        \caption{Germany: crisis}
        \label{fig:fdhd:deu:crisis}
    \end{subfigure}
    \hfill
    \begin{subfigure}[b]{0.48\textwidth}
        \centering
        \includegraphics[width=\linewidth]{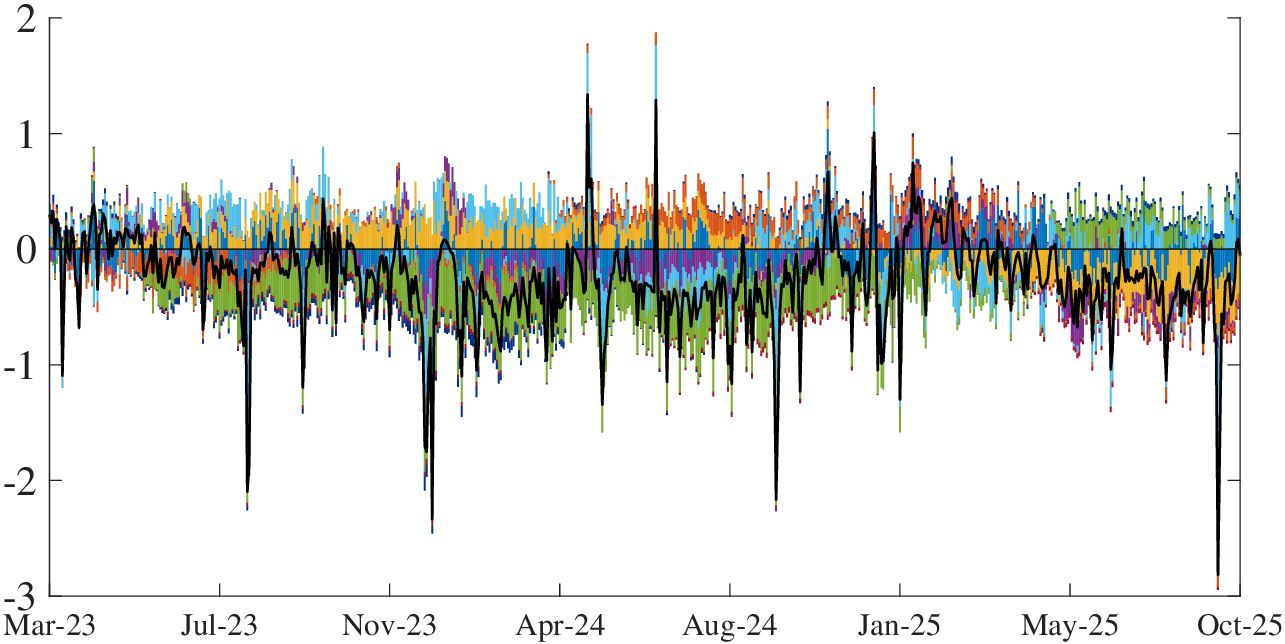}
        \caption{Germany: post-crisis}
        \label{fig:fdhd:deu:postcrisis}
    \end{subfigure}

    \medskip

    \begin{subfigure}[b]{0.48\textwidth}
        \centering
        \includegraphics[width=\linewidth]{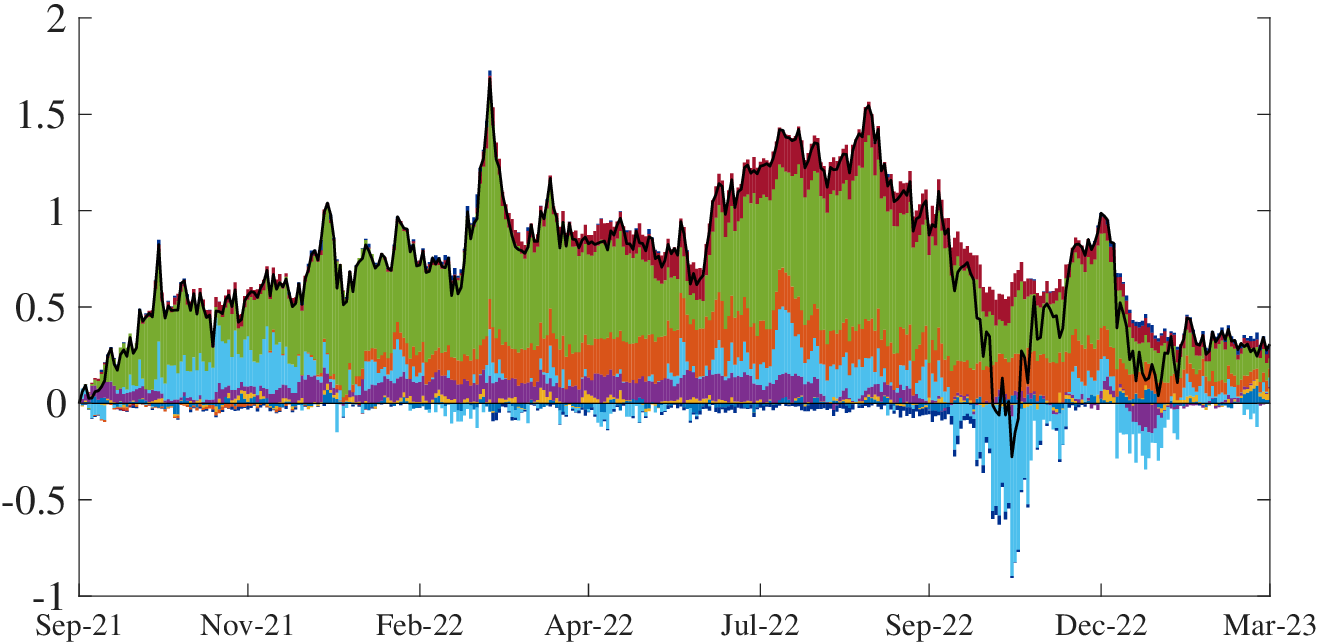}
        \caption{Italy: crisis}
        \label{fig:fdhd:ita:crisis}
    \end{subfigure}
    \hfill
    \begin{subfigure}[b]{0.48\textwidth}
        \centering
        \includegraphics[width=\linewidth]{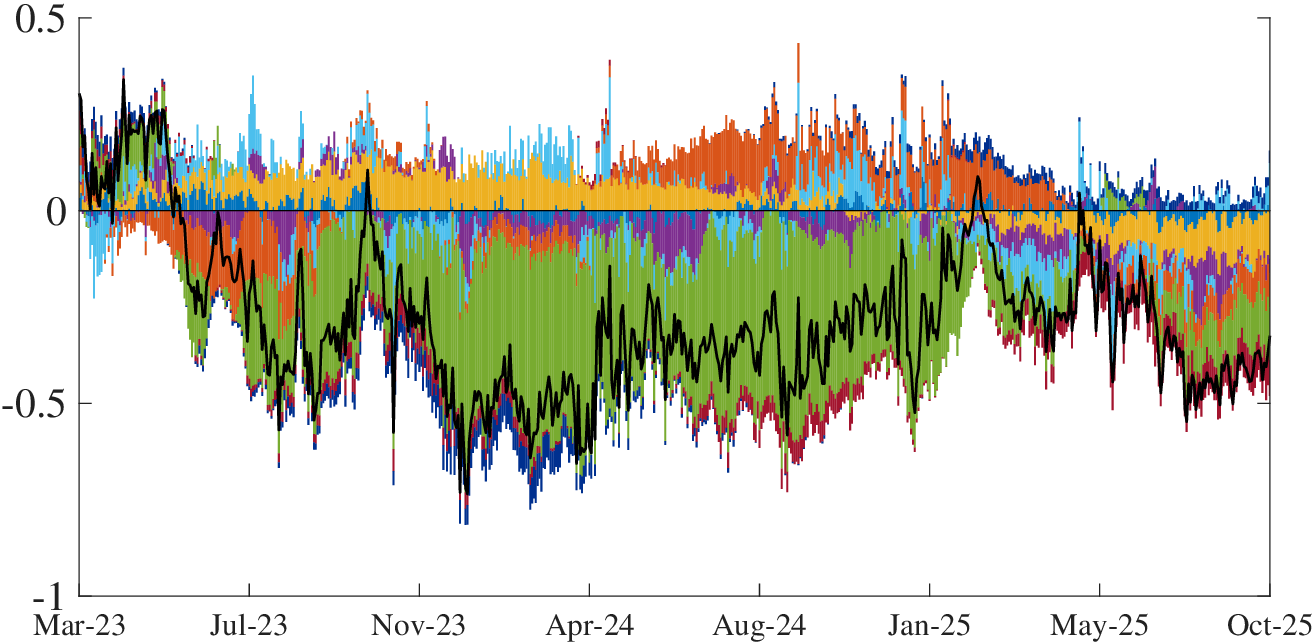}
        \caption{Italy: post-crisis}
        \label{fig:fdhd:ita:postcrisis}
    \end{subfigure}

    \caption{Historical decomposition of spot electricity prices in Germany and Italy based on the full-sample estimates. The crisis panels cover September 2021--{February} 2023 and the post-crisis panels cover March 2023--{September} 2025. The solid line displays the forecast error, defined as the deviation from the conditional forecast anchored at $T^{*}$ (see Appendix \ref{sec:app:fdhd}). The colored bars represent the cumulative contributions of structural innovations: wind (blue), solar (yellow), demand (purple), electricity (light blue), oil (orange), gas (green), coal (dark red), and CO$_2$ (navy). {All four panels are based on the same full-sample estimates and the same starting date.}}
    \label{sec:res:fig:fdhd}
\end{figure}

We conclude the presentation of the empirical results by discussing the historical decompositions of day-ahead electricity prices in Figure~\ref{sec:res:fig:fdhd}, where Panels~\ref{fig:fdhd:deu:crisis}--\ref{fig:fdhd:deu:postcrisis} refer to Germany and Panels~\ref{fig:fdhd:ita:crisis}--\ref{fig:fdhd:ita:postcrisis} to Italy. {To make the figures easier to inspect, we present each country’s decomposition in two separate plots, one for the crisis period and one for the post-crisis period.}

{Given the persistence of the endogenous variables, initial conditions may continue to influence their subsequent values. Nevertheless, we {can fix the starting date of} the historical decomposition and retain these initial conditions (see, e.g., \citealp{kilianlutkepohl2017}; Ch. 4)}. First, we choose an ``anchor'' date $T^{*}$, and then we compute recursive predictions using the full-sample VAR estimates up to horizon $h$ based on the information available on $T^{*}$. This produces a series of cumulative forecast errors which, for any horizon until $T^{*}+h$, can be decomposed into the cumulative {contributions} of each structural shock (see Appendix \ref{sec:app:fdhd} for more details).

{The decomposition starts on September~1, 2021 and continues through September~30, 2025. Therefore, it spans not only} the main phase of the European energy crisis, {but also the later developments of the post-crisis period. Notice that the smoother Italian profile reflects the use of a national price index that averages zonal prices using volume weights, dampening zone-specific fluctuations, whereas the German series refers to a single bidding zone.}

{Gas-price shocks make a large positive contribution to the forecast errors during 2022, particularly in Italy. Their contribution subsequently declines and becomes negative during 2023, remaining important over much of the post-crisis period. This reversal complements the positive gas-price responses discussed earlier: lower gas costs can provide substantial relief precisely because electricity prices remain exposed to gas. As discussed in the FEVD section, \cite{gme2024report} documents the continuing importance of gas-fired plants in price setting, even as their electricity sales decline. Oil-price shocks also make a non-negligible positive contribution during much of 2022. This need not reflect oil-fired generation alone: the gas-to-oil switching documented in European industry provides context for the indirect transmission channel discussed alongside the IRFs \citep{iea2023gasdemand}.}

{The German forecast errors display more frequent and pronounced short-run fluctuations, with substantial contributions from wind and electricity-market-specific shocks. These fluctuations can temporarily offset the relief associated with gas. On December~12, 2024, for example, the wind and electricity-market contributions are positive while the gas contribution is negative. This pattern is consistent with the account provided by \cite{acer2025key} of low wind generation, higher demand during cold weather, and substantial outages at fossil-fuel plants.}

Finally, we consider the negative contribution of electricity-market-specific shocks {during autumn 2022}. As already discussed in the introduction, this period coincided with discussions and announcements of emergency EU interventions. Anticipated reductions in electricity consumption {may have contributed} to ease price pressures by reducing the need for expensive generation, whereas the inframarginal revenue cap primarily aimed to finance consumer support. {The European Commission's subsequent review links falling electricity prices to lower gas prices, supported by demand-reduction measures alongside mild weather \citep{europeancommission2023emergencyreview}. Expectations surrounding a possible gas-price cap may also have played a role, although \citet{acer2023mcmassessment}'s subsequent assessment did not identify significant market effects clearly attributable to the adoption of the gas-market correction mechanism. While we cannot establish this link directly, the post-crisis decomposition reveals substantial negative contributions from gas-shocks, consistent with downward pressure on spot electricity prices.}

\section{Discussion and conclusions}
\label{sec:conclusion}

This paper studies the structural drivers of day-ahead electricity prices in Germany and Italy during the European energy crisis and its aftermath. Using a daily SVAR with renewable generation forecasts, load forecasts, fossil-fuel prices, carbon prices, and electricity prices, we decompose electricity-price movements into economically interpretable shocks. The results show that electricity-price dynamics cannot be understood from pairwise correlations alone: gas and electricity prices co-move strongly, but electricity-price innovations reflect the interaction of several structural shocks whose relevance varies across horizons, countries, and market conditions.

The main empirical message is twofold. First, renewable-supply shocks generate a clear merit-order effect, especially in Germany, where wind shocks account for a sizeable share of electricity-price variation and reduce day-ahead prices on impact. Second, fossil-fuel shocks become increasingly important at longer horizons, with gas-price shocks playing a central role in both countries and explaining a large part of the 2022 crisis-period surge. The comparison between Germany and Italy shows that common European shocks are transmitted through different national electricity systems: Germany displays a stronger renewable-supply channel, while Italy is more strongly driven by gas-price and electricity-market-specific shocks. 

These findings have direct policy implications {and return us to the distinction outlined in the introduction between reducing fossil-fuel dependence and managing the associated price risk.} {Diversification of gas suppliers can improve energy security by reducing dependence on individual producers, yet it does not insulate national electricity markets from global gas-price movements \citep{acer2025key}. In the longer run, expanding low-carbon generation together with flexibility could reduce the frequency with which gas-fired plants set electricity prices. The continuing importance of gas in the Italian results makes this a particularly relevant policy priority.} 

{From this perspective, Germany's closure of its remaining nuclear plants in April 2023 may have been counterproductive, removing an existing low-carbon alternative when reducing fossil-fuel exposure remained particularly important. Indeed, \citet{glynos2024nuclear} estimate that the temporary extension of nuclear operation into early 2023 reduced gas-fired generation and wholesale electricity prices relative to a phase-out at the end of 2022. Italy has instead renewed its interest in nuclear generation, adopting legislation in September 2026 to develop the regulatory framework for its return \citep{italy2026nuclear}. This represents a longer-term diversification option, however, rather than an immediate response to electricity-price pressures.}

{Alongside these longer-term choices about the generation mix, firms and consumers need ways to manage the price risk they continue to face. Hedging and long-term contracts offer complementary protection in this respect. Power Purchase Agreements and well-designed Contracts for Difference can stabilise contracted costs and revenues and support investment while preserving incentives for efficient short-run production and consumption \citep{ambec2023electricityviews, decarolis2026electricitymarketdesign}. Such arrangements do not, by themselves, remove the system's dependence on fossil fuels, but can limit the financial exposure of firms and consumers. This matters beyond the electricity sector: higher energy costs, uncertainty, and financing constraints can weaken corporate investment, particularly among energy-intensive firms \citep{anaya2025investment}.}

At the same time, renewable-supply shocks lower electricity prices mainly in the short run. This suggests that the price-stabilizing role of renewables depends not only on additional renewable capacity, but also on the flexibility required to shift renewable production across time. Battery storage is especially relevant in this respect, because it can absorb excess wind and solar generation when prices are low and release electricity when renewable output is scarce and fossil-fuel units would otherwise remain marginal. {However, expanding battery storage requires a reliable supply of critical minerals (see, e.g., \citealp{vespignani2024minerals} on how technical and non-technical risks in mineral projects can raise financing costs and discourage investment). Reducing fossil-fuel dependence therefore also requires attention to the supply chains supporting the transition}.
More broadly, storage, transmission, and flexible demand are necessary complements to renewable expansion if the merit-order benefits of renewables are to translate into more persistent reductions in electricity-price exposure. 
{Looking ahead, growing electricity demand from artificial intelligence, data centres, and industrial electrification may increase reliance on dispatchable generation and strengthen the transmission of fossil-fuel price shocks to electricity prices. However, the magnitude of these effects will depend on future investments in renewable generation, storage, and system flexibility.}

{Future research direction could incorporate macroeconomic variables, such as inflation or an industrial production index, to discuss whether the structural shocks originating in electricity markets affect economic activity. Moreover, it will be interesting to explore alternative identification strategies, such as sign restrictions or external instruments, particularly when extending the model to include macroeconomic variables.}

\pagebreak
\bibliographystyle{chicago}
\bibliography{references}

\newpage
\appendix

\section{Appendix}

\subsection{Specification of the deterministic components}
\label{sec:append_harmon}
Electricity prices, forecast demand (electricity load), and renewable wind and solar generation are characterized by high seasonality and deterministic patterns (e.g., increasing production over time). Moreover, electricity demand is subject to calendar effects given by celebrations and holidays. Finally, our sample spans the European energy crisis, which lasted from late 2021 to mid 2023. We consider this period, during which electricity and fossil prices rose sharply, as a structurally different economic regime, possibly described by a VAR with different drift terms.

Following the VAR model in Eq.\eqref{section:svar:eq:svar}, we denote the vector of deterministic terms as
\begin{equation}
    \boldsymbol{\nu}_t = \begin{pmatrix} 1, & \mathbf{h}_t^{\prime}, & \mathbf{d}_t^{\text{(cal)}^{\prime}} \end{pmatrix}^{\prime},
\end{equation}
where each component is defined as follows:
\begin{enumerate}    
    \item \textbf{Seasonal harmonics $\mathbf{h}_t$:} a vector of Fourier series to capture the smooth, deterministic seasonality inherent in renewable generation, demand, and energy prices. For a specified number of harmonics $N_h$, the vector is constructed as $\mathbf{h}_t =(\sin(\omega_1 t), \allowbreak \cos(\omega_1 t), \allowbreak \dots, \allowbreak \sin(\omega_{N_h} t), \allowbreak \cos(\omega_{N_h} t))^{\prime}$, where $\omega_j = 2\pi j / 365$ for $j=1,\ldots, N_h$.
    \item \textbf{Calendar effect dummies $\mathbf{d}_t^{\text{(cal)}}$:} electricity demand exhibits jumps around national holidays and bridge days. Because these extreme values may distort the VAR parameter estimates, we consider level-shift dummies when such an anomaly is detected (taking $1$ on the specific day $t$, and $0$ otherwise).
\end{enumerate}

\subsection{Calendar anomaly detection strategy}
We identify outliers in electricity demand by detecting observations that are ``too large'' after suitably standardizing the sample. In particular, under independently and identically normally distributed VAR errors, the sample mean and sample standard deviation are not consistent estimates in the presence of outliers, so we instead standardize observations as absolute deviations from the median, scaled by the Median Absolute Deviation (MAD).

Denoting $\widehat{u}_{it}$ as the VAR residual for series $i$, we first estimate a VAR model and then compute the ``robustly'' standardized residuals as
\begin{equation}
\label{section:app:eq:std}
    \Tilde{u}_{it} = \frac{|\widehat{u}_{it} - \text{Median}(\widehat{\textbf{u}}_{i})|}{\text{MAD}(\widehat{\textbf{u}}_{i})},
\end{equation}
where
\begin{equation*}
    \text{MAD}(\widehat{\textbf{u}}_{i}) = 1.4826 \times \text{Median}\left( |\widehat{u}_{it} - \text{Median}(\widehat{\textbf{u}}_{i})| \right).
\end{equation*}
Specifically, the factor $1.4826$ ensures that the MAD is a consistent estimate of the true standard deviation.

We therefore flag an observation as a calendar anomaly if its robustly standardized residual in Eq. \eqref{section:app:eq:std} exceeds a critical threshold, that is, if $\Tilde{u}_{it} > \tau$. In our application, we set $\tau = 4$ and condition the procedure to negative residuals. For each flagged observation, a corresponding column is added to $\mathbf{d}_t^{\text{(cal)}}$, where $d_{it}^{\text{(cal)}} = 1$ if $\tau > 4$ and zero otherwise, and finally re-estimate the VAR model.

\subsection{Historical decomposition}
\label{sec:app:fdhd}
Canonical historical decompositions rely on the existence of an infinite-order moving-average (MA) representation of the VAR. However, this representation is only valid when the true data generating process is a stable VAR. In this paper, we estimate a VAR model including time series displaying persistent autocorrelation even at relatively long lags, limiting the applicability of this decomposition tool.

Despite this limitation, we can still construct a historical decomposition starting from a meaningful ``anchor'' date $T^*$ and compute the cumulative contribution of structural innovations to realizations $\mathbf{y}_{T^*+h}$, that is, $h$ periods following the anchor date.

To derive it, we cast a reduced-form VAR($p$) process into its companion VAR(1) form:
\begin{equation*}
    Y_t = \mathbf{N}_t + \mathbf{A} Y_{t-1} + U_t,
\end{equation*}
where $Y_t =(\mathbf{y}_t^{\prime}, \mathbf{y}_{t-1}^{\prime}, \dots, \mathbf{y}_{t-p+1}^{\prime})^{\prime}$ is a $Kp \times 1$ state vector, $\mathbf{N}_t = (\boldsymbol{\nu}_t^{\prime}\mathbf{\tilde{C}}^{\prime}, 0, \dots, 0)^{\prime}$, $U_t =(\mathbf{u}_t^{\prime}, 0, \dots, 0)^{\prime}$, and $\mathbf{A}$ is the corresponding $Kp \times Kp$ companion matrix and $\mathbf{\tilde{C}} = \mathbf{B_0}^{-1} \mathbf{C}$. By recursively iterating this equation $h$ steps forward from the anchor date $T^*$, we obtain
\begin{equation*}
    Y_{T^*+h} = \sum_{i=0}^{h-1} \mathbf{A}^i \mathbf{N}_{T^*+h-i} + \mathbf{A}^h Y_{T^*} + \sum_{i=0}^{h-1} \mathbf{A}^i U_{T^*+h-i},
\end{equation*}
for any $h>0$. To extract the original $K \times 1$ vector of interest, $\mathbf{y}_{T^*+h}$, we pre-multiply the system by the $K \times Kp$ selection matrix $\mathbf{J} =(\mathbf{I}_K, 0, \dots, 0)$:
\begin{equation*}
    \mathbf{y}_{T^*+h} = \boldsymbol{\mu}_{T^*+h}^{(h)} + \mathbf{J} \mathbf{A}^h Y_{T^*} + \sum_{i=0}^{h-1} \mathbf{\Phi}_i \mathbf{u}_{T^*+h-i},
\end{equation*}
where $\boldsymbol{\mu}_{T^*+h}^{(h)} = \mathbf{J} \sum_{i=0}^{h-1} \mathbf{A}^i \mathbf{N}_{T^*+h-i}$ represents the accumulated deterministic path, and $\mathbf{\Phi}_i = \mathbf{J}\mathbf{A}^i\mathbf{J}^{\prime}$ are the reduced-form impulse response matrices. Note that $\mathbf{J}\mathbf{A}^i U_t = \mathbf{J}\mathbf{A}^i \mathbf{J}^{\prime}\mathbf{J}U_t = \mathbf{\Phi}_i \mathbf{u}_t$. The term $\boldsymbol{\mu}_{T^*+h}^{(h)} + \mathbf{J} \mathbf{A}^h Y_{T^*}$ is exactly the $h$-step-ahead baseline forecast conditional on the information set available at time $T^*$, denoted as $E(\mathbf{y}_{T^*+h} | \Omega_{T^*})$.

Subtracting this conditional baseline from the actual realization yields the $h$-step-ahead forecast error. Therefore, we can express the reduced-form errors in terms of the structural shocks as $\mathbf{u}_t = \mathbf{B}_0^{-1}\mathbf{w}_t$ and obtain
\begin{equation}
\label{Appendix:Eq:FDHD}
    \mathbf{y}_{T^*+h} - E(\mathbf{y}_{T^*+h} | \Omega_{T^*}) = \sum_{i=0}^{h-1} \mathbf{\Phi}_i \mathbf{B}_0^{-1} \mathbf{B}_0 \mathbf{u}_{T^*+h-i} = \sum_{i=0}^{h-1} \mathbf{\Theta}_i \mathbf{w}_{T^*+h-i},
\end{equation}
where $\mathbf{\Theta}_i$ are the structural impulse response matrices. According to Eq.~\eqref{Appendix:Eq:FDHD}, the $h$-step-ahead forecast error can be decomposed as the cumulative sum of structural shocks  weighted by the corresponding impulse response from time $T^*$ onward. It is therefore straightforward to recover the historical decomposition of $\mathbf{y}_{T^*+h}$.

\end{document}